\documentstyle[12pt]{article}
\begin{document}

\pagenumbering{arabic}
%\thispagestyle{empty}

%DEFINITIONS
\def\pxb{\left(\p \times \B - \B \times \p \right)}
\def\LAMBDA{\mbox{\rlap{$\raise3pt\hbox{--}$}{$\lambda$}}}
\def\rk{r_k}
\def\beq{\begin{equation}}
\def\eeq{\end{equation}}
\def\bea{\begin{eqnarray}}
\def\eea{\end{eqnarray}}
\def\nn{\nonumber}
\def\ba{\begin{array}}
\def\ea{\end{array}}
\def\0{{\mbox{\boldmath $0$}}}
\def\one{1\hskip -1mm{\rm l}}
\def\A{{\mbox{\boldmath $A$}}}
\def\B{{\mbox{\boldmath $B$}}}
\def\El{{\mbox{\boldmath $E$}}}
\def\F{{\mbox{\boldmath $F$}}}
\def\S{{\mbox{\boldmath $S$}}}
\def\P{{\mbox{\boldmath $P$}}}

\def\a{{\mbox{\boldmath $a$}}}
\def\p{{\mbox{\boldmath $p$}}}
\def\hatp{{\widehat{\mbox{\boldmath $p$}}}}
\def\vpi{{\mbox{\boldmath $\pi$}}}
\def\hatvpi{\widehat{\mbox{\boldmath $\pi$}}}
\def\r{{\mbox{\boldmath $r$}}}
\def\v{{\mbox{\boldmath $v$}}}
\def\w{{\mbox{\boldmath $w$}}}
\def\H{{\rm H}}
\def\hA{\widehat{A}}
\def\hB{\widehat{B}}
\def\ih{\frac{\i}{\hbar}}
\def\ixh{\i \hbar}
\def\ddx{\frac{\partial}{\partial x}}
\def\ddy{\frac{\partial}{\partial y}}
\def\ddz{\frac{\partial}{\partial z}}
\def\ddzn{\frac{d }{d z}}
\def\ddt{\frac{\partial}{\partial t}}
\def\vsig{{\mbox{\boldmath $\sigma$}}}
\def\Al{{\mbox{\boldmath $\alpha$}}}
\def\ho{\widehat{\cal H}_o}
\def\half{\frac{1}{2}}
\def\E{{\widehat{\cal E}}}
\def\O{{\widehat{\cal O}}}
\def\eps{\epsilon}
\def\g{\gamma}
\def\Vomeg{{\underline{\mbox{\boldmath $\Omega$}}}_s}
\def\hH{\widehat{H}}
\def\Vsig{{\mbox{\boldmath $\Sigma$}}}
\def\Nab{{\mbox{\boldmath $\nabla$}}}
\def\curl{{\rm curl}}
\def\bh{\bar{H}}
\def\th{\tilde{H}}

\def\zone{z^{(1)}}
\def\ztwo{z^{(2)}}
\def\zi{z_{\rm in}}
\def\zo{z_{\rm out}}

\def\At{{\widehat{A}(t)}}
\def\dAt{\frac{\partial {\widehat{A}(t)} }{\partial t}}
\def\sone{{\widehat{S}_1}}
\def\dsone{\frac{\partial {\widehat{S}_1} }{\partial t}}
\def\dO{\frac{\partial {\widehat{\cal O}}}{\partial t}}
\def\e{{\rm e}}
\def\ct{\widehat{\cal T}}

\begin{center}
{\LARGE\bf
An alternate way to obtain the aberration expansion
in Helmholtz Optics}

\bigskip

{\em Sameen Ahmed KHAN} \\

\bigskip

khan@fis.unam.mx,  ~ rohelakhan@yahoo.com \\
http://www.pd.infn.it/$\sim$khan/ \\
http://www.imsc.ernet.in/$\sim$jagan/khan-cv.html \\
Centro de Ciencias F\'{i}sicas, \\
Universidad Nacional Aut\'onoma de M\'exico (UNAM) \\
Apartado Postal 48-3,
Cuernavaca 62251,
Morelos, \\
{\bf M\'EXICO} \\

\end{center}

\medskip
\medskip

\begin{abstract}
Exploiting the similarities between the Helmholtz wave equation and
the Klein-Gordon equation, the former is linearized using the
Feschbach-Villars procedure used for linearizing the Klein-Gordon
equation.  Then the Foldy-Wouthuysen iterative diagonalization technique
is applied to obtain a Hamiltonian description for a system with varying
refractive index.  Besides reproducing all the traditional quasiparaxial
terms, this method leads to additional terms, which are dependent on
the wavelength, in the optical Hamiltonian.  This alternate prescription
to obtain the aberration expansion is applied to the axially symmetric
graded index fiber.  This results in the wavelength-dependent
modification of the paraxial behaviour and the aberration coefficients.
Explicit expression for the modified coefficients of the aberration to
third-order are presented.  Sixth and eighth order Hamiltonians are
derived for this system.
\end{abstract}

\def\i{{\rm i}}
%\newpage

\tableofcontents

\newpage

\setcounter{section}{0}

\section{Introduction}
The traditional scalar wave theory of optics (including aberrations to
all orders) is based on the beam-optical Hamiltonian derived using the
Fermat's principle.  This approach is purely geometrical and works
adequately in the scalar regime.  The other approach is based on the
Helmholtz equation which is derived from the Maxwell equations;  Then
one makes the {\em square-root} of the Helmholtz operator followed by
an expansion of the radical~\cite{DFW,Dragt-Wave}.  This approach
works to all orders and the resulting expansion is {\it no} different
from the one obtained using the geometrical approach of the Fermat's
principle.

Another way of obtaining the aberration expansion is based on the
algebraic similarities between the Helmholtz equation and the
Klein-Gordon equation.  Exploiting this algebraic similarity the
Helmholtz equation is linearized in a procedure very similar to the one
due to Feschbach-Villars, for linearizing the  Klein-Gordon equation.
This brings the Helmholtz equation to a Dirac-like form and then
follows the procedure of the Foldy-Wouthuysen expansion used in the
Dirac electron theory.  This  approach, which uses the algebraic
machinery of quantum mechanics, was developed recently~\cite{KJS-1},
providing an alternative to the traditional {\em square-root}
procedure.  This scalar formalism gives rise to wavelength-dependent
contributions modifying the aberration coefficients.  The algebraic
machinery of this formalism is very similar to the one used in the
{\em quantum theory of charged-particle beam optics}, based on the
Dirac~\cite{JSSM}-\cite{J0} and the Klein-Gordon~\cite{KJ1} equations
respectively.  The detailed account for both of these is available
in~\cite{JK2}.  A treatment of beam optics taking into account the
anomalous magnetic moment is available in~\cite{CJKP-1}-\cite{JK}.

General expressions for the Hamiltonians are derived without assuming
any specific form for the refractive index.  These Hamiltonians are
shown to contain the extra wavelength-dependent contributions which
arise very naturally in our approach.  We apply the general formalism
to the specific examples:
A. {\em Medium with Constant Refractive Index}.  This example
is essentially for illustrating some of the details of the machinery
used.
The Feschbach-Villars technique for linearizing the Klein-Gordon
equation is summarized in Appendix-A.  The Foldy-Wouthuysen
transformation technique is outlined in Appendix-B.

The other application, B. {\em Axially Symmetric Graded Index Medium}
is used to demonstrate the power of the formalism.  The traditional
approaches give six aberrations.  Our formalism modifies these six
aberration coefficients by wavelength-dependent contributions.

The traditional beam-optics is completely obtained from our approach
in the limit wavelength, $\LAMBDA \longrightarrow 0$, which we call as
the traditional limit of our formalism.  This is analogous to the
classical limit obtained by taking $\hbar \longrightarrow 0$ in the
quantum prescriptions.  The scheme of using the Foldy-Wouthuysen
machinery in this formalism is very similar to the one used in the
{\em quantum theory of charged-particle beam
optics}~\cite{JSSM}-\cite{JK}.  There  too one recovers the classical
prescriptions in the limit $\lambda_0 \longrightarrow 0$ where
$\lambda_0 = {\hbar}/{p_0}$ is the de Broglie wavelength and $p_0$ is
the design momentum of the system under study.

\section{Traditional Prescriptions}
Recalling, that in the traditional scalar wave theory for treating
monochromatic quasiparaxial light beam propagating along the positive
$z$-axis, the $z$-evolution of the optical wave function $\psi(\r)$ is
taken to obey the Schr\"{o}dinger-like equation
%01
\bea
\i \LAMBDA\frac{\partial }{\partial z} \psi (\r)
= \widehat{H} \psi (\r)\,,
\label{Schr}
\eea
where the optical Hamiltonian $\widehat{H}$ is formally given by the
radical
%02
\bea
\widehat{H} = - \left({n^2 (\r) - \hatp_\perp^2} \right)^{1/2}\,,
\eea
and the refractive index, $n (\r) = n (x , y , z)$.  In beam optics the
rays are assumed to propagate almost parallel to the optic-axis, chosen
to be $z$-axis, here.  That is,
$\left| \hatp_\perp \right| \ll p_z \approx 1$ and
$\left| n (\r) - n_0 \right| \ll n_0$.
The refractive index is the order of unity.  Let us further assume that
the refractive index varies smoothly around the constant background
value $n_0$ without any abrupt jumps or discontinuities.    For a medium
with uniform refractive index, $n (\r) = n_0$ and
the Taylor expansion of the radical is
%03
\bea
\left({n^2 (\r) - \hatp_\perp^2} \right)^{1/2}
& = &
n_0 \left\{1 - \frac{1}{n_0^2} \hatp_\perp^2 \right\}^{1/2} \nn \\
& = &
n_0 \left\{
1 - \frac{1}{2 n_0^2} \hatp_\perp^2
- \frac{1}{8 n_0^4} \hatp_\perp^4
- \frac{1}{16 n_0^6} \hatp_\perp^6 \right. \nn \\
& & \left. \quad \qquad \qquad
- \frac{5}{128 n_0^8} \hatp_\perp^8
- \frac{7}{256 n_0^{10}} \hatp_\perp^{10} - \cdots
\right\}\,.
\label{Taylor-Expansion}
\eea
In the above expansion one retains terms to any desired degree of
accuracy in powers of
$\left(\frac{1}{n_0^2} \hatp_\perp^2\right)$.  In general the
refractive index is not a constant and varies.  The variation of the
refractive index $n (\r)$, is expressed as a Taylor expansion in the
spatial variables $x$, $y$ with $z$-dependent coefficients.  To get
the beam optical Hamiltonian one makes the expansion of the radical
as before, and retains terms to the desired order of accuracy in
$\left(\frac{1}{n_0^2} \hatp_\perp^2\right)$ along with all the other
terms (coming from the expansion of the refractive index $n (\r)$) in
the phase-space components up to the same order.  In this expansion
procedure the problem is partitioned into paraxial behaviour $+$
aberrations, order-by-order.

In relativistic quantum mechanics too, one has the problem of
understanding the behaviour in terms of nonrelativistic limit $+$
relativistic corrections, order-by-order.  In the Dirac theory of the
electron this is done most conveniently through the Foldy-Wouthuysen
transformation.

Here, we follow a procedure similar to the one used for linearizing
the Klein-Gordon equation {\em via} the Feshbach-Villars linearizing
procedure~\cite{FV}.  The resulting Feshbach-Villars-like form has an
algebraic structure very similar to the Dirac equation.  This enables
us to make an expansion using the Foldy-Wouthuysen transformation
technique well-known in the Dirac electron theory~\cite{Foldy,BD}.
The resulting expansion reproduces the above expansion
in~(\ref{Taylor-Expansion}) as it should.  Furthermore it gives rise to
a set of wavelength-dependent contributions.  The formalism presented
here is an elaboration of the recent work which provides an alternative
to the traditional square-root technique of obtaining the optical
Hamiltonian~\cite{KJS-1}.

Let us start with the wave-equation in the rectilinear coordinate
system.
%04
\bea
\left\{\Nab^2 - \frac{n^2 (\r)}{v^{2}}
\frac{\partial^2 }{\partial t^2} \right\} \Psi = 0\,.
\label{wave-equation}
\eea
Let
%05
\bea
\Psi = \psi (\r) e^{- \i \omega t}\,, \qquad \qquad
\omega > 0\,,
\label{time-gone}
\eea
then
%06
\bea
\left\{\Nab^2 + \frac{n^2 (\r)}{v^2} \omega^2 \right\}
\psi (\r) = 0\,.
\label{Helmholtz}
\eea
At this stage we introduce the wavization,
%07
\bea
- \i \LAMBDA \Nab_\perp \longrightarrow \hatp_\perp\,,
\qquad \qquad
- \i \LAMBDA \frac{\partial }{\partial z}
\longrightarrow p_z\,,
\label{wavization-1}
\eea
where $\LAMBDA$ is the reduced wavelength given by
$\LAMBDA = {\lambda}/{2 \pi}$, $c = \LAMBDA \omega$ and
$n (\r) = {c}/{v (\r)}$.
It is to be noted that $pq -qp = - \i \LAMBDA$.  This is similar
to the commutation relation in quantum mechanics.
In our formalism $\LAMBDA$ plays the same role which is played by
the Planck constant, $\hbar$ in quantum mechanics.  The traditional
beam-optics formalism is completely obtained from our formalism in
the limit $\LAMBDA \longrightarrow 0$.
Then, we get,
%08
\bea
\left\{
\left(- \i \LAMBDA \frac{\partial }{\partial z} \right)^2
+
\left(\hatp_{\perp}^2 - n^2 (\r) \right)
\right\} \psi (\r) = 0\,.
\label{wavization-2}
\eea

Next, we linearize Eq.~(\ref{wavization-2}) following a procedure
similar to, the one which gives the Feshbach-Villars~\cite{FV} form of
the Klein-Gordon equation.  To this end, let
%09
\bea
\left(
\ba{c}
\psi_1 (\r) \\
\psi_2 (\r) \\
\ea
\right)
=
\left(
\ba{c}
\psi (\r) \\
- \i \frac{\LAMBDA}{n_0} \ddz \psi (\r)
\ea
\right)\,.
\eea
Then, Eq.~(\ref{wavization-2}) is equivalent to
%10
\bea
- \i \frac{\LAMBDA}{n_0} \ddz
\left(
\ba{c}
\psi_1 (\r) \\
\psi_2 (\r) \\
\ea
\right) %\nn \\
=
\left(
\ba{ccc}
0 & & 1 \\
\frac{1}{n_0^2}\left(n^2 (\r) - \hatp_{\perp}^2 \right)
& & 0 \\
\ea
\right)
\left(
\ba{c}
\psi_1 (\r) \\
\psi_2 (\r) \\
\ea
\right)
\label{Equivalent}
\eea
Next, we make the transformation,
%11
\bea
\left(
\ba{c}
\psi_1 \left(\r \right) \\
\psi_2 \left(\r \right) \\
\ea
\right)
\longrightarrow
\Psi^{(1)}
& = &
\left(
\ba{c}
\psi_{+} \left(\r \right) \\
\psi_{-} \left(\r \right) \\
\ea
\right)
=
M
\left(
\ba{c}
\psi_1 (\r) \\
\psi_2 (\r) \\
\ea
\right) \nn \\
& = &
\frac{1}{\sqrt{2}}
\left(
\ba{c}
\psi_1 (\r) + \psi_2 (\r) \\
\psi_1 (\r) - \psi_2 (\r) \\
\ea
\right) \nn \\
& = &
\frac{1}{\sqrt{2}}
\left(
\ba{c}
\psi (\r)
- \i \frac{\LAMBDA}{n_0} \ddz \psi (\r) \\
\psi (\r)
+ \i \frac{\LAMBDA}{n_0} \ddz \psi (\r) \\
\ea
\right)
\label{Transformation}
\eea
where
%12
\bea
M = M^{-1} =
\frac{1}{\sqrt{2}}
\left(
\ba{lr}
1 & 1 \\
1 & -1
\ea
\right)\,, \qquad
\det M = - 1\,.
\label{M}
\eea
It is to be noted that the transformation matrix $M$ is independent
of $z$.  For a monochromatic quasiparaxial beam (in forward direction),
with leading $z$-dependence
$\psi (\r) \sim \exp{\{{\i n (\r) z}/{\LAMBDA}\}}$.  Then
%13
\bea
\psi_{+}
& \sim &
\frac{1}{\sqrt{2}}
\left\{1 + \frac{n (\r)}{n_0} \right\} \psi (\r) \nn \\
\psi_{-}
& \sim &
\frac{1}{\sqrt{2}}
\left\{1 - \frac{n (\r)}{n_0} \right\} \psi (\r)
\eea
Since, $\left| n (\r) - n_0 \right| \ll n_0$, we have
$\psi_{+} \gg \psi_{-}$.

Consequently, Eq.~(\ref{wavization-2}) can be written as
%14
\bea
\i \LAMBDA
\ddz
\left(
\ba{c}
\psi_{+} (\r) \\
\psi_{-} (\r)
\ea
\right)
& = &
\widehat{\H}
\left(
\ba{c}
\psi_{+} (\r) \\
\psi_{-} (\r)
\ea
\right)\,, \nn \\
\widehat{\H}
& = &
- n_0 \sigma_z + \widehat{\cal E}
+ \widehat{\cal O} \nn \\
\widehat{\cal E} & = & \frac{1}{2 n_0}
\left\{\hatp_{\perp}^2 + \left(n_0^2 - n^2 \left(\r \right) \right)
\right\} \sigma_z \nn \\
\widehat{\cal O}
& = &
\frac{1}{2 n_0}
\left\{\hatp_{\perp}^2 + \left(n_0^2 - n^2 (\r) \right) \right\}
\left(\i \sigma_y \right)\,,
\label{FV-FW}
\eea
where $\sigma_y$ and $\sigma_z$ are, respectively, the $y$ and
$z$ components of the triplet of Pauli matrices,
%15
\beq
\vsig = \left(
\sigma_x =
\left(
\ba{cc}
0 & 1 \\
1 & 0
\ea
\right), \ \
\sigma_y =
\left(
\ba{lr}
0 & -\i \\
\i & 0
\ea \right), \ \
\sigma_z =
\left(
\ba{lr}
1 & 0 \\
0 & -1
\ea
\right)
\right).
\eeq
It is to be noted that the even-part and odd-part in
Hamiltonian~(\ref{FV-FW}) differ only by a Pauli matrix.  This
simplifies the commutations a lot as we shall see, shortly.
The details of the Feshbach-Villars linearizing procedure for the
Klein-Gordon equation are available in Appendix-A.

The square of the Hamiltonian is
%16
\bea
\widehat{\H}^2
& = &
\left\{\left(n^2 \left(\r \right)
- \hatp_{\perp}^2 \right) \right\}\,,
\eea
as expected.  Thus we have have taken the square-root is a different
way.  This has certain distinct advantages over the traditional
procedure of directly taking the square-root.

The purpose of casting Eq.~(\ref{wavization-2}) in the form of
Eq.~(\ref{FV-FW}) will be obvious now, when we compare the latter
with the form of the Dirac equation
%17
\bea
\i \hbar \frac{\partial}{\partial t}
\left(
\ba{c}
\Psi_u \\
\Psi_l
\ea
\right)
& = &
\widehat{H}_D
\left(
\ba{c}
\Psi_u \\
\Psi_l
\ea
\right) \nn \\
\widehat{H}_D
& = &
{m_0 c^2} \beta + \widehat{\cal E}_D + \widehat{\cal O}_D \nn \\
\widehat{\cal E}_D
& = &
q \phi \nn \\
\widehat{\cal O}_D
& = &
c \Al \cdot \widehat{\vpi}\,,
\label{Full-Dirac}
\eea
where $u$ and $l$ stand for the upper and lower components
respectively and
%18
\bea
{\mbox{\boldmath $\alpha$}}
& = &
\left[
\ba{cc}
{\mbox{\boldmath $0$}} & {\mbox{\boldmath $\sigma$}} \\
{\mbox{\boldmath $\sigma$}} & {\mbox{\boldmath $0$}}
\ea
\right]\,, \qquad
\beta
=
\left[
\ba{cc}
\one & {\mbox{\boldmath $0$}} \\
{\mbox{\boldmath $0$}} & - \one
\ea
\right]\,, \qquad
\one
=
\left[
\ba{cc}
1 & 0 \\
0 & 1
\ea
\right]\,.
\eea

To proceed further, we note the striking similarities between
Eq.~(\ref{FV-FW}) and Eq.~(\ref{Full-Dirac}).  In the nonrelativistic
positive energy case, the upper components $\Psi_u$ are large
compared to the lower components $\Psi_l$.  The odd ($\widehat{\cal O}$)
part of ($\widehat{H}_D - {m_0 c^2} \beta$), anticommuting with $\beta$
couples the large $\Psi_u$ to $\Psi_l$ while the even
($\widehat{\cal E}$) part commuting with $\beta$, does not couple them.
Using this fact, the well known Foldy-Wouthuysen formalism of the Dirac
electron theory (see, {\it e.g.},~\cite{BD}) employs a series of
transformations on Eq.~(\ref{Full-Dirac}) to reach a representation in
which the Hamiltonian is a sum of the nonrelativistic part and a series
of relativistic correction terms;
${\left| c \widehat{\vpi} \right|}/{m_0 c^2}$ serves as the
expansion parameter and the nonrelativistic part corresponds to an
approximation of order up to
${\left| c \widehat{\vpi} \right|}/{m_0 c^2}$.  The terms of higher
order in ${\left| c \widehat{\vpi} \right|}/{m_0 c^2}$ constitute the
relativistic corrections.  Examining Eq.~(\ref{FV-FW}) we conclude
$\psi_{+} \gg \psi_{-}$, and the odd operator $\widehat{\cal O}$,
anticommuting with $\sigma_z$, couples the large $\psi_{+}$ with
the small $\psi_{-}$, while the even operator $\widehat{\cal E}$ does
not make such a coupling.  This spontaneously suggests that a
Foldy-Wouthuysen-{\it like} technique can be used to transform
Eq.~(\ref{FV-FW}) into a representation in which the corresponding
beam optical Hamiltonian is a series with expansion parameter
${\left| \hatp_{\perp} \right|}/{n_0}$.  The correspondence between
the beam optical Hamiltonian~(\ref{FV-FW}) and the Dirac electron
theory is summarized in the following table:

%\bigskip

\newpage

\begin{center}
{\bf The Analogy}
\end{center}
\begin{tabular}{ll}
{\bf Standard Dirac Equation} ~~~~~~~ & {\bf Beam Optical Form} \\
$m_0 c^2 \beta + \E_D + \O_D$ & $- n_0 \sigma_z + \E + \O$ \\
$m_0 c^2$ & $- n_0$ \\
Positive Energy & Forward Propagation \\
Nonrelativistic, $\left| \widehat{\vpi} \right| \ll m_0 c$ &
Paraxial Beam, $\left|\hatp_\perp \right| \ll n_0$ \\
%Deviations from nonrelativistic situation & Deviations from paraxial condition \\
Non relativistic Motion  & Paraxial Behavior \\
~~ + Relativistic Corrections & ~~ + Aberration Corrections \\
\end{tabular}

\bigskip

Application of the Foldy-Wouthuysen-{\it like} technique to
Eq.~(\ref{FV-FW}) involves a series of transformations on it and
after the required number of transformations, depending on the
degree of accuracy, Eq.~(\ref{FV-FW}) is transformed into a form in
which the residual odd part can be neglected and hence the upper and
lower components ($\psi_{+}$ and $\psi_{-}$) are effectively decoupled.
In this representation the larger component ($\psi_{+}$) corresponds
to the beam moving in the $+ z$-direction and the smaller component
($\psi_{-}$) corresponds to the backward moving component of the beam.

Using the correspondence between Eq.~(\ref{FV-FW}) and
Eq.~(\ref{Full-Dirac}) the Foldy-Wouthuysen expansion given formally
in terms of $\E$ and $\O$ leads to the Hamiltonian
%19
\bea
\i \LAMBDA \ddz \left| \psi \right\rangle
& = &
\widehat{\cal H}^{(2)} \left|\psi \right\rangle\,, \nn \\
\widehat{\cal H}^{(2)} & = &
- n_0 \sigma_z + \E - \frac{1}{2 n_0} \sigma_z \O^2\,,
\label{FW-2-Formal}
\eea
To simplify the formal Hamiltonian we use,
$\O^2 = - \frac{1}{4 n_0^2}
\left\{\hatp_\perp^2 + \left(n_0^2 - n^2 (\r) \right) \right\}^2$
and recall that
$\E = \frac{1}{2 n_0}
\left\{\hatp_\perp^2 + \left(n_0^2 - n^2 (\r) \right) \right\}
\sigma_z$.
Dropping the $\sigma_z$ the formal Hamiltonian in
in~(\ref{FW-2-Formal}) is expressed in terms of the
phase-space variables as:
%20
\bea
\widehat{\cal H}^{(2)}
& = &
- n_0
+ \frac{1}{2 n_0}
\left\{\hatp_\perp^2 + \left(n_0^2 - n^2 (\r) \right) \right\} \nn \\
& &
+ \frac{1}{8 n_0^3}
\left\{\hatp_\perp^2 + \left(n_0^2 - n^2 (\r) \right) \right\}^2\,.
\label{H-Two}
\eea
The details of the Foldy-Wouthuysen iterative procedure are described
in detain in Appendix-B.
The lowest order Hamiltonian obtained in this procedure
agrees with  the traditional approaches.

To go beyond the expansions in~(\ref{H-Two}) one goes a step
further in the Foldy-Wouthuysen iterative procedure.
To next-to-leading order the Hamiltonian is formally given by
%21
\bea
\i \LAMBDA \ddz \left|\psi \right\rangle
& = &
\widehat{\cal H}^{(4)} \left|\psi \right\rangle\,, \nn \\
\widehat{\cal H}^{(4)}
& = &
- n_0 \sigma_z + \E - \frac{1}{2 n_0} \sigma_z \O^2 \nn \\
& & - \frac{1}{8 n_0^2}
\left[\O ,
\left(\left[\O , \E \right] + \i \LAMBDA \ddz \O \right) \right] \nn \\
& &
+ \frac{1}{8 n_0^3} \sigma_z
\left\{
\O^4
+ \left(\left[\O , \E \right] + \i \LAMBDA \ddz \O \right)^2
\right\}\,.
\label{FW-4-Formal}
\eea
As before we drop the $\sigma_z$ and the resulting Hamiltonian
in the phase-space variable is
%22
\bea
\widehat{\cal H}^{(4)}
& = &
- n_0
+ \frac{1}{2 n_0} \left\{\hatp_{\perp}^2
+ \left(n_0^2 - n^2 (\r) \right) \right\} \nn \\
& & \quad
+ \frac{1}{8 n_0^3}
\left\{\hatp_\perp^2 + \left(n_0^2 - n^2 (\r) \right) \right\}^2 \nn \\
& & \quad
- \frac{\i \LAMBDA}{32 n_0^4}
\left[\hatp_\perp^2 , \ddz \left(n^2 (\r) \right) \right] \nn \\
& & \quad
+ \frac{\LAMBDA^2}{32 n_0^5}
\left(\ddz \left(n^2 (\r) \right) \right)^2 \nn \\
& & \quad
+ \frac{1}{16 n_0^5}
\left\{\hatp_\perp^2 + \left(n_0^2 - n^2 (\r) \right) \right\}^3 \nn \\
& & \quad
+ \frac{5}{128 n_0^7}
\left\{\hatp_\perp^2 + \left(n_0^2 - n^2 (\r) \right) \right\}^4\,.
\label{H-Four}
\eea
The Hamiltonian thus derived has all the terms which one gets in
the traditional square-root approaches.  In addition we also get the
wavelength-dependent contributions.

The details of the various transforms and the beam optical formalism
being discussed here turns out to be a simplified analog of the more
general formalism recently developed for the {\em quantum theory of
charged-particle beam optics}~\cite{JSSM}-\cite{JK}, both in the
scalar and the spinor cases, respectively.  A very detailed description
of these transforms and techniques is available in~\cite{JK2}.

Now, we can compare the above Hamiltonians with the conventional
Hamiltonian given by the square-root approach~\cite{Dragt-Wave}.  The
square-root approach does not give all the terms, such as the one
involving the commutator
of $\p_\perp^2$ with $\ddz \left(n^2 (\r) \right)$.  Our procedure of
linearization and expansion in powers of
${\left| \hatp_\perp \right|}/{n_0}$ gives all the terms which one
gets by the square-root expansion of~(\ref{Taylor-Expansion})
and some additional terms, which are the wavelength-dependent terms.
Such, wavelength-dependent terms can in no way be obtained by any of
the conventional prescriptions, starting with the Helmholtz
equation~(\ref{Helmholtz}).

\section{Applications}
In the previous sections we presented an alternative to the
square-root expansion and and we obtained an expansion for the
beam-optical Hamiltonian which works to all orders.  Formal expressions
were obtained for the paraxial Hamiltonian and the leading order
aberrating Hamiltonian, without assuming any form for the refractive
index.  Even at the paraxial level the wavelength-dependent effects
manifest by the presence of a commutator term, which does not vanish
for a varying refractive index.

Now, we apply the formalism to specific examples.  First one is the
medium with constant refractive index.  This is perhaps the only
problem which can be solved exactly in a closed form expression.  This
is just to illustrate how the aberration expansion in our formalism can
be summed to give the familiar exact result.

The next example is that of the axially symmetric graded index medium.
This example enables us to demonstrate the power of the formalism,
reproducing the familiar results from the traditional approaches and
further giving rise to new results, dependent on the wavelength.

%CONSTANT
\subsection{Medium with Constant Refractive Index}
For a medium with constant refractive index,
$n (\r) = n_c$, we have,
%23
\bea
\widehat{\H}_{c}
& = &
- n_0 \sigma_z + {\cal D} \sigma_z
+ {\cal D} \left(\i \sigma_y \right) \nn \\
{\cal D}
& = &
\frac{1}{2 n_0}
\left\{\hatp_{\perp}^2 + \left(n_0^2 - n_c^2 \right) \right\}\,.
\label{H-Constant}
\eea
The Hamiltonian in~(\ref{H-Constant}) can be exactly diagonalized
by the following transform,
%24
\bea
T^{\pm}
& = &
\exp{\left[\i \left(\pm \i \sigma_z \right) \O \theta \right]} \nn \\
& = &
\exp{ \left[\mp \sigma_x {\cal D} \theta \right]} \nn \\
& = &
\cosh \left({\cal D} \theta \right)
\mp \sigma_x \sinh \left({\cal D} \theta \right)\,.
\label{T-D}
\eea
We choose,
%25
\bea
\tanh \left(2 {\cal D} \theta \right)
=
\frac{{\cal D}}{n_0 - {\cal D}}
=
\frac{n_0^2 - \left(n_c^2 - \hatp_{\perp}^2 \right)}
{n_0^2 + \left(n_c^2 + \hatp_{\perp}^2 \right)} < 1\,,
\label{tanh}
\eea
then we obtain,
%26
\bea
\widehat{\H}_c^{\rm diagonal}\,
& = &
T^{+} \widehat{\H}_{c} T^{-} \nn \\
& = &
T^{+} \left\{- n_0 \sigma_z + {\cal D} \sigma_z
+ {\cal D} \left(\i \sigma_y \right) \right\} T^{-} \nn \\
& = &
- \sigma_z \left\{n_0^2 - 2 n_0 {\cal D} \right\}^{\frac{1}{2}} \nn \\
& = &
- \sigma_z \left\{n_c^{2} - \hatp_{\perp}^2 \right\}^{\frac{1}{2}}
\label{Constant-Diagonal}
\eea
We next, compare the exact result thus obtained with the approximate
one, obtained through the systematic series procedure we have developed.
We define
$P =
\frac{1}{n_0^2} \left\{\hatp_{\perp}^2
+ \left(n_0^2 - n_c^{2} \right) \right\}$. Then,
%
%27
\bea
\widehat{\cal H}^{(4)}_c
& = &
- n_0
\left\{1 - \frac{1}{2} P - \frac{1}{8} P^2
- \frac{1}{16} P^3 - \frac{5}{128} P^4 \right\} \sigma_z \nn \\
& \approx &
- n_0 \left\{1 - P^2 \right\}^{\frac{1}{2}} \nn \\
& = &
- \left\{n_c^{2} - \hatp_{\perp}^2 \right\}^{\frac{1}{2}} \nn \\
& = &
\widehat{\H}_c^{\rm diagonal}\,.
\label{Constant-approximate}
\eea

Knowing the Hamiltonian, we can compute the transfer maps.  The
transfer operator between any pair of points
$\left\{(z^{\prime \prime} , z^{\prime}) \left|
z^{\prime \prime} \right. > z^{\prime} \right\}$
on the $z$-axis, is formally given by
%28
\bea
\left|\psi (z^{\prime \prime} , z^{\prime}) \right|
=
\widehat{\cal T} (z^{\prime \prime} , z^{\prime})
\left|\psi (z^{\prime \prime} , z^{\prime}) \right\rangle \,,
\label{}
\eea
with
%29
\bea
& & \i \LAMBDA \frac{\partial}{\partial z}
\widehat{\cal T} (z^{\prime \prime} , z^{\prime})
=
\widehat{\cal H} \widehat{\cal T} (z^{\prime \prime} , z^{\prime})\,,
\quad
\widehat{\cal T} (z^{\prime \prime} , z^{\prime})
=
\widehat{\cal I}\,, \nn \\
& & \nn \\
& & \widehat{\cal T} (z^{\prime \prime} , z^{\prime})
=
\wp \left\{\exp
\left[- \frac{\i}{\LAMBDA}
\int_{z^\prime}^{z^{\prime \prime}} dz\,
\widehat{\cal H} (z) \right] \right\} \nn \\
& & \quad
=
\widehat{\cal I}
- \frac{\i}{\LAMBDA} \int_{z^\prime}^{z^{\prime \prime}} dz
\widehat{\cal H} (z) \nn \\
& & \qquad
+ \left(- \frac{\i}{\LAMBDA}\right)^2
\int_{z^\prime}^{z^{\prime \prime}} dz
\int_{z^\prime}^{z} d z^\prime
\widehat{\cal H} (z) \widehat{\cal H} (z^\prime) \nn \\
& & \qquad
+ \ldots\,,
\label{Transfer-1}
\eea
where $\widehat{\cal I}$ is the identity operator and $\wp$ denotes the
path-ordered exponential.  There is no closed form expression for
$\widehat{\cal T} (z^{\prime \prime} , z^{\prime})$ for an arbitrary
choice of the refractive index $n (\r)$.  In such a situation the most
convenient form of the expression for the $z$-evolution
operator $\widehat{\cal T} (z^{\prime \prime} , z^{\prime})$, or the
$z$-propagator, is
%30
\beq
\widehat{\cal T} (z^{\prime \prime} , z^{\prime})
=
\exp{
\left[- \frac{\i}{\LAMBDA} \widehat{T}
(z^{\prime \prime} , z^{\prime}) \right]}\,,
\label{Transfer-2}
\eeq
with
%31
\bea
\hat {T} (z^{\prime \prime} , z^{\prime})
%& & \quad
& = &
\int_{z^\prime}^{z^{\prime \prime}} dz \widehat{\cal H} (z) \nn \\
& & \qquad
+ \frac{1}{2} \left(- \frac{\i}{\LAMBDA} \right)
\int_{z^\prime}^{z^{\prime \prime}} dz
\int_{z^\prime}^{z}  d z^\prime
\left[\widehat{\cal H} (z)\,, \widehat{\cal H} (z^\prime) \right] \nn \\
& & \qquad
+ \ldots \,,
\label{T-Magnus}
\eea
as given by the Magnus formula~\cite{Magnus} which is described in
Appendix-C.  We shall be needing these expressions in the next example
where the refractive index is not a constant.

Using the procedure outlined above we compute the transfer operator,
%32
\bea
& &
\widehat{U}_c \left(\zo\,, \zi \right)
=
\exp{ \left[- \frac{\i}{\LAMBDA} \Delta z {\cal H}_c \right] }
\nn \\
& & =
\exp{ \left[+ \frac{\i}{\LAMBDA} n_c \Delta z
\left\{1 - \frac{1}{2} \frac{\widehat{p}_{\perp}^2}{n_c^2}
- \frac{1}{8} \left(\frac{\widehat{p}_{\perp}^2}{n_c^2} \right)^2
- \cdots \right\} \right]}\,, \nn \\
& & \qquad \qquad \qquad \Delta z = \zo - \zi \,,
\label{Constant-U}
\eea
Using~(\ref{Constant-U}), we compute the transfer maps
%33
\bea
\left(
\ba{c}
\left\langle \r_{\perp} \right\rangle \\
\left\langle \p_{\perp} \right\rangle
\ea
\right)_{\rm out}
=
\left(
\ba{ccc}
1 & & \frac{1}{\sqrt{n_c^2 - \p_{\perp}^2}} \Delta z \\
0 & & 1
\ea
\right)
\left(
\ba{c}
\left\langle \r_{\perp} \right\rangle \\
\left\langle \p_{\perp} \right\rangle
\ea
\right)_{\rm in}\,.
\label{Constant-Maps}
\eea
The beam-optical Hamiltonian is intrinsically aberrating.  Even for
simplest situation of a constant refractive index, we have aberrations
to all orders!

%AXIAL
\subsection{Axially Symmetric Graded Index Medium}
The refractive index of an axially symmetric graded-index
material can be most generally described by the following
polynomial (see, pp.~117 in~\cite{DFW})
%34
\bea
n (\r) = n_0 + \alpha_2 (z) \r_{\perp}^2
+ \alpha_4 (z) \r_{\perp}^4 + \alpha_6 (z) \r_{\perp}^6 +
\alpha_8 (z) \r_{\perp}^8 + \cdots\,,
\label{n-Dragt}
\eea
where, we have assumed the axis of symmetry to coincide with the
optic-axis, namely the $z$-axis without any loss of generality.
To write the beam-optical Hamiltonians we introduce the following
notation
%35
\bea
\widehat{T}
& = &
\left(\hatp_\perp \cdot \r_\perp + \r_\perp \cdot \hatp_\perp \right)
\nn \\
w_1 (z)
& = &
\ddzn \left\{2 n_0 \alpha_2 (z) \right\} \nn \\
w_2 (z)
& = &
\ddzn \left\{\alpha_2^2 (z) + 2 n_0 \alpha_4 (z) \right\} \nn \\
w_3 (z)
& = &
\ddzn \left\{2 n_0 \alpha_6 (z) + 2 \alpha_2 (z) \alpha_4 (z) \right\} \nn \\
w_4 (z)
& = &
\ddzn \left\{\alpha_4^2 (z) + 2 \alpha_2 (z) \alpha_6 (z)
+ 2 n_0 \alpha_8 (z) \right\}
\eea

\noindent
We also use,
$\left[A , B \right]_{+} = \left(AB + BA \right)$.
The beam-optical Hamiltonian is
%36 TO
\bea
\widehat{\cal H}
& = &
\widehat{H}_{0\,, p} + \widehat{H}_{0\,, (4)}
+ \widehat{H}_{0\,, (6)} + \widehat{H}_{0\,, (8)} \nn \\
& & \quad
+ \widehat{H}_{0\,, (2)}^{(\LAMBDA)}
+ \widehat{H}_{0\,, (4)}^{(\LAMBDA)}
+ \widehat{H}_{0\,, (6)}^{(\LAMBDA)}
+ \widehat{H}_{0\,, (8)}^{(\LAMBDA)} \nn \\
\widehat{H}_{0\,, p}
& = &
- n_0
+ \frac{1}{2 n_0} \hatp_\perp^2 - \alpha_2 (z) \r_\perp^2 \nn \\
\widehat{H}_{0\,, (4)}
& = &
\frac{1}{8 n_0^3} \hatp_\perp^4
- \frac{\alpha_2 (z)}{4 n_0^2}
\left(\hatp_\perp^2 \r_\perp^2 + \r_\perp^2 \hatp_\perp^2 \right)
- \alpha_4 (z) \r_\perp^4 \nn \\
\widehat{H}_{0\,, (6)}
& = &
\frac{1}{16 n_0^5} \hatp_\perp^6
- \frac{\alpha_2 (z)}{8 n_0^4}
\left\{
\left(\hatp_\perp^4 \r_\perp^2 + \r_\perp^2 \hatp_\perp^4 \right)
+ \hatp_\perp^2 \r_\perp^2 \hatp_\perp^2 \right\} \nn \\
& & \quad
+
\frac{1}{8 n_0^3} \left\{
\left(\alpha_2^2 (z) - 2 n_0 \alpha_4 (z) \right)
\left(\hatp_\perp^2 \r_\perp^4 + \r_\perp^4 \hatp_\perp^2 \right)
+ 2 \alpha_2^2 (z) \r_\perp^2 \hatp_\perp^2 \r_\perp^2 \right\} \nn \\
& & \quad
- \alpha_6 (z) \r_\perp^6 \nn \\
\widehat{H}_{0\,, (8)}
& = &
\frac{5}{128 n_0^7} \hatp_\perp^8
- \frac{5 \alpha_2 (z)}{64 n_0^6}
\left[\hatp_\perp^4 , \left[\hatp_\perp^2  \r_\perp^2 \right]_{+}
\right]_{+} \nn \\
& & \quad
+ \frac{1}{32 n_0^5}
\left\{
\left(3 \alpha_2^2 (z) - 4 n_0 \alpha_4 (z) \right)
\left[\hatp_\perp^4 , \r_\perp^4 \right]_{+}
+
5 \alpha_2^2 (z)
\left[\hatp_\perp^2 , \r_\perp^2 \right]_{+}^2 \right. \nn \\
& & \left. \qquad \qquad \qquad
\vphantom{\left[\hatp_\perp^2 , \r_\perp^2 \right]_{+}^2}
- \left(2 \alpha_2^2 (z) + 4 n_0 \alpha_4 (z) \right)
\hatp_\perp^2 \r_\perp^4 \hatp_\perp^2
\right\} \nn \\
& & \quad
+ \frac{1}{16 n_0^4}
\left\{
4 \left(\alpha_2^3 (z) + n_0 \alpha_2 (z) \alpha_4 (z)
+ n_0^2 \alpha_6 (z) \right)
\left[\hatp_\perp^2 , \r_\perp^6 \right]_{+} \right. \nn \\
& & \left. \qquad \qquad \qquad
\vphantom{\left[\hatp_\perp^2 , \r_\perp^2 \right]_{+}^2}
- 5 \alpha_2^3 (z)
\left[\r_\perp^4 , \left[\hatp_\perp^2 , \r_\perp^2 \right]_{+}
\right]_{+} \right. \nn \\
& & \left. \qquad \qquad \qquad
\vphantom{\left[\hatp_\perp^2 , \r_\perp^2 \right]_{+}^2}
+
\left(2 \alpha_2^3 (z) + 4 n_0 \alpha_2 (z) \alpha_4 (z) \right)
\left[\r_\perp^2 , \r_\perp^2 \hatp_\perp^2 \r_\perp^2 \right]_{+}
\right\} \nn \\
& & \quad
- \alpha_8 (z) \r_\perp^8 \nn \\
\widehat{H}_{0\,, (2)}^{(\LAMBDA)}
& = &
- \frac{\LAMBDA^2}{16 n_0^4}
\left\{\ddzn \left(n_0 \alpha_2 (z) \right) \right\} \widehat{T} \nn \\
\widehat{H}_{0\,, (4)}^{(\LAMBDA)}
& = &
- \frac{\LAMBDA^2}{32 n_0^4} w_2 (z)
\left(\r_\perp^2 \widehat{T} + \widehat{T} \r_\perp^2 \right)
+
\frac{\LAMBDA^2}{32 n_0^5} w_1^2 (z) \r_\perp^4 \nn \\
\widehat{H}_{0\,, (6)}^{(\LAMBDA)}
& = &
- \frac{3 \LAMBDA^2}{32 n_0^4} w_3 (z)
\left(\r_\perp^4 \widehat{T} + \widehat{T} \r_\perp^4 \right)
+
\frac{\LAMBDA^2}{16 n_0^5} w_1 (z) w_2 (z) \r_\perp^6 \nn \\
\widehat{H}_{0\,, (8)}^{(\LAMBDA)}
& = &
- \frac{\LAMBDA^2}{8 n_0^4} w_4 (z)
\left(\r_\perp^6 \widehat{T} + \widehat{T} \r_\perp^6 \right)
+
\frac{\LAMBDA^2}{32 n_0^5} \left\{
w_2^2 (z) + 2 w_1 (z) w_3 (z) \right\} \r_\perp^8
\label{HAMILTONIAN}
\eea
The reason for partitioning $\widehat{\cal H}$ in the above manner
will be clear as we proceed.

The paraxial transfer maps are formally given by
%37
\bea
\left(
\ba{c}
\left\langle \r_{\perp} \right\rangle \\
\left\langle \p_{\perp} \right\rangle
\ea
\right)_{\rm out}
=
\left(
\ba{cc}
P & Q \\
R & S
\ea
\right)
\left(
\ba{c}
\left\langle \r_{\perp} \right\rangle \\
\left\langle \p_{\perp} \right\rangle
\ea
\right)_{\rm in}\,,
\label{Paraxial-Maps}
\eea
where~$P$, $Q$, $R$ and $S$ are the solutions of the paraxial
Hamiltonian in~(\ref{HAMILTONIAN}).

The transfer operator is most accurately expressed in terms of
the the paraxial solutions, $P$, $Q$, $R$ and $S$,
{\em via} the {\em interaction picture}~\cite{Interaction-Picture}.
%38
\bea
\widehat{\cal T} \left(z\,, z_0\right)
& = &
\exp {
\left[
- \frac{\i}{\LAMBDA} \widehat{T}
\left(z\,, z_0\right) \right] }\,, \nn \\
& = &
\exp
\left[
- \frac{\i}{\LAMBDA}
\left\{
C \left(z^{\prime \prime}\,, z^\prime \right) \hatp_{\perp}^4
\phantom{\frac{\i}{\LAMBDA}} \right. \right. \nn \\
& & \qquad \qquad \quad
+
K \left(z^{\prime \prime}\,, z^\prime \right)
\left[\hatp_{\perp}^2 \,, \left(\hatp_\perp \cdot \r_\perp
+ \r_\perp \cdot \hatp_\perp \right) \right]_{+} \nn \\
& & \qquad \qquad \quad
+ A \left(z^{\prime \prime}\,, z^\prime \right)
\left(\hatp_\perp \cdot \r_\perp
+ \r_\perp \cdot \hatp_\perp \right)^{2} \nn \\
& & \qquad \qquad \quad
+
F \left(z^{\prime \prime}\,, z^\prime \right)
\left(\hatp_{\perp}^2 \r_{\perp}^2
+ \r_{\perp}^2 \hatp_{\perp}^2 \right) \nn \\
& & \qquad \qquad \quad
+
D \left(z^{\prime \prime}\,, z^\prime \right)
\left[\r_{\perp}^2 \,, \left(\hatp_\perp \cdot \r_\perp
+ \r_\perp \cdot \hatp_\perp \right) \right]_{+} \nn \\
& & \qquad \qquad \quad \left. \left.
+
E \left(z^{\prime \prime}\,, z^\prime \right)
\r_{\perp}^4
\vphantom{\frac{\i}{\LAMBDA}}
\right\} \right]\,,
\eea

The six aberration coefficients are given by,
%39
\bea
C \left(z^{\prime \prime}\,, z^\prime \right)
& = &
\int_{z^\prime}^{z^{\prime \prime}} d z
\left\{
\frac{1}{8 n_0^3} S^4
- \frac{\alpha_2 (z)}{2 n_0^2} Q^2 S^2
- \alpha_4 (z) Q^4 \right. \nn \\
& & \left. \qquad \qquad \qquad \quad
- \frac{\LAMBDA^2}{8 n_0^4} w_2 (z) Q^3 S
+ \frac{\LAMBDA^2}{32 n_0^5} w_1^2 (z) Q^4
\right\} \nn \\
K \left(z^{\prime \prime}\,, z^\prime \right)
& = &
\int_{z^\prime}^{z^{\prime \prime}} d z
\left\{
\frac{1}{8 n_0^3} R S^3
- \frac{\alpha_2 (z)}{4 n_0^2} QS(PS + QR)
- \alpha_4 (z) PQ^3 \right. \nn \\
& & \left. \qquad \qquad \qquad \quad
- \frac{\LAMBDA^2}{32 n_0^4} w_2 (z)
\left(Q^2 (PS + QR) + 2 P Q^2 S \right) \right. \nn \\
& & \left. \qquad \qquad \qquad \quad
+ \frac{\LAMBDA^2}{32 n_0^5} w_1^2 (z) P Q^3
\right\} \nn \\
A \left(z^{\prime \prime}\,, z^\prime \right)
& = &
\int_{z^\prime}^{z^{\prime \prime}} d z
\left\{
\frac{1}{8 n_0^3} R^2 S^2
- \frac{\alpha_2 (z)}{2 n_0^2} PQRS
- \alpha_4 (z) P^2 Q^2 \right. \nn \\
& & \left. \qquad \qquad \qquad \quad
- \frac{\LAMBDA^2}{16 n_0^4} w_2 (z)
\left(PQ (PS + QR) \right) \right. \nn \\
& & \left. \qquad \qquad \qquad \quad
+ \frac{\LAMBDA^2}{32 n_0^5} w_1^2 (z) P^2 Q^2
\right\} \nn \\
F \left(z^{\prime \prime}\,, z^\prime \right)
& = &
\int_{z^\prime}^{z^{\prime \prime}} d z
\left\{
\frac{1}{8 n_0^3} R^2 S^2
- \frac{\alpha_2 (z)}{4 n_0^2} (P^2 S^2 + Q^2 R^2)
- \alpha_4 (z) P^2 Q^2 \right. \nn \\
& & \left. \qquad \qquad \qquad \quad
- \frac{\LAMBDA^2}{16 n_0^4} w_2 (z)
\left(PQ (PS + QR) \right) \right. \nn \\
& & \left. \qquad \qquad \qquad \quad
+ \frac{\LAMBDA^2}{32 n_0^5} w_1^2 (z) P^2 Q^2
\right\} \nn \\
D \left(z^{\prime \prime}\,, z^\prime \right)
& = &
\int_{z^\prime}^{z^{\prime \prime}} d z
\left\{
\frac{1}{8 n_0^3} R^3 S
- \frac{\alpha_2 (z)}{4 n_0^2} PR (PS + QR)
- \alpha_4 (z) P^3 Q \right. \nn \\
& & \left. \qquad \qquad \qquad \quad
- \frac{\LAMBDA^2}{32 n_0^4} w_2 (z)
\left(P^2 (PS + QR) + 2 P^2 Q R \right) \right. \nn \\
& & \left. \qquad \qquad \qquad \quad
+ \frac{\LAMBDA^2}{32 n_0^5} w_1^2 (z) P^3 Q
\right\} \nn \\
E \left(z^{\prime \prime}\,, z^\prime \right)
& = &
\int_{z^\prime}^{z^{\prime \prime}} d z
\left\{
\frac{1}{8 n_0^3} R^4
- \frac{\alpha_2 (z)}{2 n_0^2} P^2 R^2
- \alpha_4 (z) P^4 \right. \nn \\
& & \left. \qquad \qquad \qquad \quad
- \frac{\LAMBDA^2}{8 n_0^4} w_2 (z)
\left(P^3 R \right) \right. \nn \\
& & \left. \qquad \qquad \qquad \quad
+ \frac{\LAMBDA^2}{32 n_0^5} w_1^2 (z) P^4
\right\}\,.
\label{fiber-aberration-coefficients}
\eea

Thus we see that the transfer operator and the aberration coefficients
are modified by $\LAMBDA$-dependent contributions.

The sixth and eighth order Hamiltonians are modified by the
presence of wavelength-dependent terms.  These will in turn modify the
the fifth and seventh order aberrations
respectively~\cite{KBW-1}-\cite{KBW-4}.

%CONCLUDING
\section{Concluding Remarks}
We exploited the similarities between the Helmholtz equation and the
Klein-Gordon equation to obtain an alternate prescription for the
aberration expansion.  In this prescription we followed a procedure
due to Feschbach-Villars for linearizing the Klein-Gordon equation.
After casting the Helmholtz equation to this linear form, it was
further possible to use the Foldy-Wouthuysen transformation technique
of the Dirac electron theory.  This enabled us to obtain the
beam-optical Hamiltonian to any desired degree of accuracy.  We further
get the wavelength-dependent contributions to at each order, starting
with the lowest-order paraxial paraxial Hamiltonian.  Formal
expressions were obtained for the paraxial and leading order aberrating
Hamiltonians, without making any assumption on the form of the
refractive index.

As an example we considered the {\em medium with a constant refractive
index}.  This is perhaps the only problem which can be solved exactly,
in a closed form expression.  This example was primarily for
illustrating certain aspects of the machinery we have used.

The second, and the more interesting example is that of the
{\em axially symmetric graded index medium}.  For this system we
derived the beam-optical Hamiltonians to eighth order.  At each order
we find the wavelength-dependent contributions.  The fourth order
Hamiltonian was used to obtain the six, third order aberrations
coefficients which get modified by the wavelength-dependent
contributions.  Explicit relations for these coefficients were
presented.  In the limit $\LAMBDA \longrightarrow 0$, the alternate
prescription here, reproduces the very well known {\it Lie Algebraic
Formalism of Light Optics}.  It would be worthwhile to look for the
extra wavelength-dependent contributions experimentally.

The close analogy between geometrical optics and charged-particle
has been known for too long a time.  Until recently it was possible
to see this analogy only between the geometrical optics and classical
prescriptions of charge-particle optics.  A quantum theory of
charged-particle optics was presented in recent
years~\cite{JSSM}-\cite{JK}.  With the  current development of the
non-traditional prescriptions of Helmholtz optics~\cite{KJS-1}
and the matrix formulation of Maxwell optics, using the rich algebraic
machinery of quantum mechanics, it is now possible to see a parallel of
the analogy at each level.  The non-traditional prescription of the
Helmholtz optics is in close analogy with the quantum theory of
charged-particles based on the Klein-Gordon equation.  The matrix
formulation of Maxwell optics presented here is in close analogy with
the quantum theory of charged-particles based on the Dirac
equation~\cite{Analogy-ICFA}.  We shall examine the parallel of these
analogies in Appendix-D and summarize the Hamiltonians in the various
prescriptions in Table-A.

\newpage

\newpage
\setcounter{section}{0}
%\setcounter{page}{1}
%\pagenumbering{arabic}

\section*{}
\addcontentsline{toc}{section}
{Appendix A. \\
The Feshbach-Villars Form of the Klein-Gordon Equation}

%FV
\renewcommand{\theequation}{A.{\arabic{equation}}}
\setcounter{equation}{0}

\begin{center}

{\Large\bf
Appendix A. \\
The Feshbach-Villars Form of the Klein-Gordon Equation
} \\

\end{center}

The method we have followed to cast the time-independent
Klein-Gordon equation into a beam optical form  linear in
$\frac{\partial}{\partial z}$, suitable for a systematic study,
through successive approximations, using the Foldy-Wouthuysen-like
transformation technique borrowed from the Dirac theory, is
similar to the way the time-dependent Klein-Gordon equation is
transformed (Feshbach and Villars,~\cite{FV}) to
the Schr\"{o}-dinger form,
containing only first-order time derivative, in order to study its
nonrelativistic limit using the Foldy-Wouthuysen technique
(see, {\em e.g.}, Bjorken and Drell,~\cite{BD}).

Defining
%01
\beq
\Phi = \frac{\partial}{\partial t} \Psi \,,
\eeq
the free particle Klein-Gordon equation is written as
%02
\beq
\frac{\partial}{\partial t} \Phi =
\left( c^2 {\nabla}^2 - \frac{ {m_0}^2 c^4}{ {\hbar}^2} \right) \Psi \,.
\eeq
Introducing the linear combinations
%03
\beq
\Psi _{+} = \half \left( \Psi + \frac{\i \hbar}{m_0 c^2} \Phi \right)\,,
\qquad
\Psi _{-} = \half \left( \Psi - \frac{\i \hbar}{m_0 c^2} \Phi \right)
\eeq
the Klein-Gordon equation is seen to be equivalent to a pair of of
coupled  differential equations:
%04
\bea
\i \hbar \frac{\partial}{\partial t} \Psi _{+} & = &
- \frac{ {\hbar}^2 {\nabla}^2 }{ 2 m_0} \left( \Psi _{+} +
\Psi _{-} \right) + m_0 c^2 \Psi _{+}
\nn \\
\i \hbar \frac{\partial}{\partial t} \Psi _{-} & = &
\frac{ {\hbar}^2 {\nabla}^2 }{ 2 m_0} \left( \Psi _{+} +
\Psi _{-} \right) - m_0 c^2 \Psi _{-}\,.
\label{PM}
\eea
Equation~(\ref{PM}) can be written in a two-component language as
%05
\beq
\i \hbar \frac{\partial}{\partial t}
{\left(
\begin{array}{c}
\Psi _{+} \\
\Psi _{-}
\end{array}
\right)}
=
\widehat{H}_0^{FV}
{\left(
\begin{array}{c}
\Psi _{+} \\
\Psi _{-}
\end{array}
\right)}\,,
\eeq
with the Feshbach-Villars Hamiltonian for the free particle,
$\widehat{H}_0^{FV}$,  given by
%06
\bea
\widehat{H}_0^{FV} & = &
{\left(
\begin{array}{cc}
m_0 c^2 + \frac{\widehat{p}^2}{2 m_0} &
\frac{\widehat{p}^2}{2 m_0} \\
- \frac{\widehat{p}^2}{2 m_0} &
- m_0 c^2 - \frac{\widehat{p}^2}{2 m_0} \\
\end{array}
\right)} \nn \\
& = &
m_0 c^2 \sigma_z + \frac{\widehat{p}^2}{2 m_0} \sigma_z +
\i \frac{\widehat{p}^2}{2 m_0} \sigma_y\,.
\eea
For a free nonrelativistic particle with kinetic energy $\ll m_0 c^2$,
it is seen that $\Psi _{+}$ is large compared to $\Psi _{-}$.

In presence of an electromagnetic field, the interaction is introduced
through the minimal coupling
%07
\beq
\widehat{\p} \longrightarrow
\widehat{\vpi} = \widehat{\p} - q \A \,, \qquad
\i \hbar \frac{\partial}{\partial t} \longrightarrow
\i \hbar \frac{\partial}{\partial t} - q \phi.
\eeq
The corresponding Feshbach-Villars form of the Klein-Gordon equation
becomes
%08
\bea
\i \hbar \frac{\partial}{\partial t}
{\left(
\begin{array}{c}
\Psi_{+} \\
\Psi_{-}
\end{array}
\right)}
& = &
\widehat{H}^{FV}
{\left(
\begin{array}{c}
\Psi_{+} \\
\Psi_{-}
\end{array}
\right)} \nn \\
{\left(
\begin{array}{c}
\Psi_{+} \\
   \\
\Psi_{-}
\end{array}
\right)}
& = &  \half
{\left(
\begin{array}{c}
\Psi + \frac{1}{m_0 c^2} \left(\i \hbar \frac{\partial}{\partial t}
- q \phi \right) \Psi \\
    \\
\Psi - \frac{1}{m_0 c^2} \left(\i \hbar \frac{\partial}{\partial t}
- q \phi \right) \Psi \\
\end{array}
\right)}
\nn \\
\widehat{H}^{FV} & = &
m_0 c^2 \sigma_z + \widehat{\cal E} + \widehat{\cal O} \nn \\
\widehat{\cal E}
& = &  q \phi + \frac{\widehat{\pi}^2}{2 m_0} \sigma_z\,, \quad
\widehat{\cal O}
=
i \frac{\widehat{\pi}^2}{2 m_0} \sigma_y\,.
\label{FV-KG}
\eea
As in the free-particle case, in the nonrelativistic situation
$\Psi_{+}$ is large compared to $\Psi_{-}$. The even term
$\widehat{\cal E}$ does not couple $\Psi_{+}$ and $\Psi_{-}$ whereas
$\widehat{\cal O}$ is odd which couples $\Psi_{+}$ and $\Psi_{-}$.
Starting from~(\ref{FV-KG}), the nonrelativistic limit of the
Klein-Gordon equation, with various correction terms, can be
understood using the Foldy-Wouthuysen technique (see,~{\em e.g.},
Bjorken and Drell,~\cite{BD}).

It is clear from the above that we have just adopted the above
technique for studying the $z$-evolution of the Klein-Gordon
wavefunction of a charged-particle beam in an optical system comprising
a static electromagnetic field.  The additional feature of our
formalism is the extra approximation of dropping $\sigma_z$ in an
intermediate stage to take into account the fact that we are interested
only in the forward-propagating beam along the $z$-direction.

\newpage

\section*{}
\addcontentsline{toc}{section}
{Appendix B. \\
The Foldy-Wouthusysen Representation of the Dirac Equation}

%FW
\renewcommand{\theequation}{B.{\arabic{equation}}}
\setcounter{equation}{0}

\begin{center}

{\Large\bf
Appendix-B. \\
Foldy-Wouthuysen Transformation
} \\

\end{center}

In the traditional scheme the purpose of expanding the {\em light
optics} Hamiltonian
$\widehat{H}
= - \left(n^2 (\r) - \hatp_\perp^2\right)^{1/2}$ in a
series using $\left(\frac{1}{n_0^2} \hatp_\perp^2\right)$ as the
expansion parameter is to understand the propagation of the
quasiparaxial beam in terms of a series of approximations
(paraxial + nonparaxial).  Similar is the situation in the case of the
{\em charged-particle optics}.  Let us recall that in relativistic
quantum mechanics too one has a similar problem of understanding
the relativistic wave equations as the nonrelativistic approximation
plus the relativistic correction terms in the quasirelativistic regime.
For the Dirac equation (which is first order in time) this is done most
conveniently using the Foldy-Wouthuysen transformation leading to an
iterative diagonalization technique.

The main framework of the formalism of optics, used here (and in the
charged-particle optics) is based on the transformation technique of
the Foldy-Wouthuysen theory which casts the Dirac equation in a form
displaying the different interaction terms between the Dirac particle
and and an applied electromagnetic field in a nonrelativistic and
easily interpretable form (see,~\cite{Foldy,Acharya}, for a general
discussion of the role of the Foldy-Wouthuysen-type transformations in
particle interpretation of relativistic wave equations).  In the
Foldy-Wouthuysen theory the Dirac equation is decoupled through a
canonical transformation into two two-component equations: one reduces
to the Pauli equation in the nonrelativistic limit and the other
describes the negative-energy states.

Let us describe here briefly the standard Foldy-Wouthuysen theory so
that the way it has been adopted for the purposes of the above studies
in optics will be clear.  Let us consider a charged-particle of
rest-mass $m_0$, charge $q$ in the presence of an electromagnetic field
characterized by
$\El = - \Nab \phi
- \frac{\partial }{\partial t} {\mbox{\boldmath $A$}}$
and $\B = \Nab \times {\mbox{\boldmath $A$}}$.
Then the Dirac equation is
%01-02
\bea
\i \hbar \frac{\partial}{\partial t} \Psi(\r , t)
& = &
\widehat{H}_D \Psi(\r , t)
\label{A-FW-1} \\
\widehat{H}_D
& = &
{m_0 c^2} \beta + q \phi + c \Al \cdot \widehat{\vpi} \nn \\
& = &
{m_0 c^2} \beta + \widehat{\cal E} + \widehat{\cal O} \nn \\
\widehat{\cal E}
& = &
q \phi \nn \\
\widehat{\cal O}
& = &
c \Al \cdot \widehat{\vpi}\,,
\label{A-FW-2}
\eea
where
%03
\bea
{\mbox{\boldmath $\alpha$}}
& = &
\left[
\ba{cc}
{\mbox{\boldmath $0$}} & {\mbox{\boldmath $\sigma$}} \nn \\
{\mbox{\boldmath $\sigma$}} & {\mbox{\boldmath $0$}}
\ea
\right]\,, \qquad
\beta
=
\left[
\ba{cc}
\one & {\mbox{\boldmath $0$}} \nn \\
{\mbox{\boldmath $0$}} & - \one
\ea
\right]\,, \qquad
\one
=
\left[
\ba{cc}
1 & 0 \nn \\
0 & 1
\ea
\right]\,, \nn \\
{\mbox{\boldmath $\sigma$}}
& = &
\left[
\sigma_x =
\left[
\ba{cc}
0 & 1 \\
1 & 0
\ea
\right]\,, \
\sigma_y =
\left[
\ba{lr}
0 & - \i \\
\i & 0
\ea
\right]\,, \
\sigma_z =
\left[
\ba{lr}
1 & 0 \\
0 & -1
\ea
\right]
\right].
\eea
with
$
\widehat{\vpi}
=
{\widehat{\mbox{\boldmath $p$}}} - q {\mbox{\boldmath $A$}}$,
$\widehat{\mbox{\boldmath $p$}} = - \i \hbar \Nab$, and
$\widehat{\pi}^2 =
\left(\widehat{\pi}_x^2 + \widehat{\pi}_y^2
+ \widehat{\pi}_z^2\right)$.

In the nonrelativistic situation the upper pair of components of the
Dirac Spinor $\Psi$ are large compared to the lower pair of components.
The operator $\widehat{\cal E}$ which does not couple the large and
small components of $\Psi$ is called `even' and $\widehat{\cal O}$ is
called an `odd' operator which couples the large to the small
components.  Note that
%04
\beq
\beta \widehat{\cal O} = - \widehat{\cal O} \beta\,, \qquad
\beta \widehat{\cal E} = \widehat{\cal E} \beta\,.
\eeq
Now, the search is for a unitary transformation,
$\Psi'$ $=$ $\Psi$ $\longrightarrow$ $\widehat{U} \Psi$, such that the
equation for $\Psi '$ does not contain any odd operator.

In the free particle case (with $\phi = 0$
and $\widehat{\vpi} = \widehat{\p}$)
such a Foldy-Wouthuysen transformation is given by
%05
\bea
\Psi \longrightarrow \Psi'
& = &
\widehat{U}_{F} \Psi \nn \\
\widehat{U}_F
& = & e^{\i \widehat{S}} =
e^{\beta \Al \cdot \widehat{\p} \theta}\,,
\quad {\rm tan}\,2 |\widehat{\p}|
\theta = \frac{| \widehat{\p}|}{m_0 c}\,.
\eea
This transformation eliminates the odd part completely from the
free particle Dirac Hamiltonian reducing it to the diagonal form:
%06
\bea
\i \hbar \frac{\partial}{\partial t} \Psi'
& = &
e^{\i \widehat{S}} \left({m_0 c^2} \beta
+ c \Al \cdot \widehat{\p} \right)
e^{- \i \widehat{S}} \Psi ' \nn \\
& = &
\left(\cos \,| \widehat{\p}| \theta +
\frac{\beta \Al \cdot \widehat{\p}}{| \widehat{\p} |} \sin \,|
\widehat{\p} | \theta
\right)
\left({m_0 c^2} \beta + c \Al \cdot \widehat{\p} \right) \nn \\
& & \qquad \qquad
\times \left(\cos \,| \widehat{\p}| \theta -
\frac{\beta \Al \cdot \widehat{\p}}{| \widehat{\p} |} \sin \,|
\widehat{\p} | \theta
\right) \Psi' \nn \\
& = &
\left(m_0 c^2 \cos \,2 | \widehat{\p}| \theta + c
|\widehat{\p}| \sin\, 2 |\widehat{\p}| \theta \right)
\beta \Psi' \nn \\
& = &
\left(\sqrt{m_0^2 c^4 + c^2 \widehat{p}^2} \right) \beta \,\Psi'\,.
\eea

In the general case, when the electron is in a time-dependent
electromagnetic field it is not possible to construct an
$\exp (\i \widehat{S})$ which removes the odd operators from the
transformed Hamiltonian completely. Therefore, one has to be content
with a nonrelativistic expansion of the transformed Hamiltonian in a
power series in $1/{m_0 c^2}$ keeping through any desired order.
Note that in the nonrelativistic case, when $|\p| \ll m_0 c$,
the transformation operator $\widehat{U}_F = \exp (\i \widehat{S})$
with $\widehat{S} \approx - \i \beta \widehat{\cal O}/{2 m_0 c^2}$,
where
$\widehat{\cal O} = c \Al \cdot \widehat{\p}$ is the odd part of the
free Hamiltonian.  So, in the general case we can start with the
transformation
%07
\beq
\Psi^{(1)} = e^{\i \widehat{S}_1} \Psi, \qquad \widehat{S}_1 =
- \frac{\i \beta \widehat{\cal O}}{2 m_0 c^2}
= - \frac{\i \beta \Al \cdot \widehat{\vpi }}{2 m_0 c}\,.
\eeq
Then, the equation for $\Psi^{(1)}$ is
%08
\bea
\i \hbar \frac{\partial}{\partial t} \Psi^{(1)}
& = &
\i \hbar \frac{\partial}{\partial t} \left(e^{\i \widehat{S}_1}
\Psi \right)
=
\i \hbar \frac{\partial}{\partial t} \left(e^{\i \widehat{S}_1}
\right) \Psi + e^{\i \widehat{S}_1} \left(\i \hbar
\frac{\partial}{\partial t} \Psi\right) \nn \\
& = &
\left[\i \hbar \frac{\partial}{\partial t}
\left(e^{\i \widehat{S}_1} \right) + e^{\i \widehat{S}_1}
\widehat{H}_D \right] \Psi \nn \\
& = &
\left[\i \hbar \frac{\partial}{\partial t}
\left(e^{\i \widehat{S}_1} \right)
e^{- \i \widehat{S}_1}
+ e^{\i \widehat{S}_1} \widehat{H}_D e^{- \i \widehat{S}_1}
\right] \Psi^{(1)} \nn \\
& = &
\left[
e^{\i \widehat{S}_1} \widehat{H}_D e^{- \i \widehat{S}_1}
- \i \hbar e^{\i \widehat{S}_1}
\frac{\partial}{\partial t} \left(e^{- \i \widehat{S}_1} \right)
\right] \Psi^{(1)} \nn \\
& = &
\widehat{H}_D^{(1)} \Psi^{(1)}
\eea
where we have used the identity $\frac{\partial}{\partial t}
\left(e^{\widehat{A}} \right) e^{- \widehat{A}}$ $+$ $e^{\widehat A}
\frac{\partial}{\partial t} \left(e^{- \widehat{A}} \right)$
$=$ $\frac{\partial}{\partial t} \widehat{I}$ $= 0$.

Now, using the identities
%09
\bea
e^{\widehat{A}} \widehat{B} e^{- \widehat{A}}
& = &
\widehat{B} + [\widehat{A} , \widehat{B} ]
+ \frac{1}{2!} [\widehat{A} , [\widehat{A} , \widehat{B} ]]
+ \frac{1}{3!} [\widehat{A} , [\widehat{A} , [\widehat{A} ,
\widehat{B} ]]] +
\ldots \nn \\
& & e^{\widehat{A}(t)} \frac{\partial}{\partial t}
\left(e^{- \widehat{A}(t)} \right) \nn \\
& & \ \
= \left( 1 + {\widehat{A}(t)} + \frac{1}{2!} {\widehat{A}(t)}^2
+ \frac{1}{3!} {\widehat{A}(t)}^3 \cdots \right) \nn \\
& & \ \
\quad \quad \times \frac{\partial}{\partial t}
\left(1 - {\widehat{A}(t)} + \frac{1}{2!} {\widehat{A}(t)}^2
- \frac{1}{3!} {\widehat{A}(t)}^3 \cdots \right) \nn \\
& & \ \
= \left(1 + \At + \frac{1}{2!} \At^2
+ \frac{1}{3!} \At^3 \cdots \right) \nn \\
& & \ \ \quad \quad
\times \left(- \dAt + \frac{1}{2!} \left\{\dAt \At +
\At \dAt \right\} \right. \nn \\
& & \ \ \quad \quad
- \frac{1}{3!} \left\{\dAt \At^2 + \At \dAt \At \right.  \nn \\
& & \ \ \quad \quad \left. \left.
+ \At^2 \dAt \right\} \ldots \right) \nn \\
& & \ \ \approx
- \dAt - \frac{1}{2!} \left[\At , \dAt \right] \nn \\
& & \ \ \quad \quad
- \frac{1}{3!} \left[\At , \left[\At , \dAt \right] \right] \nn \\
& & \ \ \quad \quad
- \frac{1}{4!} \left[\At , \left[ \At , \left[ \At , \dAt
\right] \right] \right]\,,
\eea
with $\widehat{A} = {\i \widehat{S}_1}$, we find
%10
\bea
\widehat{H}_D^{(1)}
& \approx &
\widehat{H}_D - \hbar \dsone
+ \i \left[\sone , \widehat{H}_D - \frac{\hbar}{2} \dsone \right] \nn \\
& & \qquad
- \frac{1}{2!} \left[ \sone , \left[\sone ,
\widehat{H}_D - \frac{\hbar}{3} \dsone \right] \right] \nn \\
& & \qquad
- \frac{\i}{3!} \left[ \sone , \left[\sone , \left[\sone ,
\widehat{H}_D - \frac{\hbar}{4} \dsone \right] \right] \right]\,.
\label{A-FW-9}
\eea
Substituting in~(\ref{A-FW-9}),
$\widehat{H}_D = {m_0 c^2} \beta + \widehat{\cal E}
+ \widehat{\cal O}$, simplifying the right hand side using the
relations $\beta \widehat{\cal O} = - \widehat{\cal O} \beta$
and $\beta \widehat{\cal E} = \widehat{\cal E} \beta$ and collecting
everything together, we have
%11
\bea
\widehat{H}_D^{(1)}
& \approx &
{m_0 c^2} \beta + \widehat{\cal E}_1 + \widehat{\cal O}_1 \nn \\
\widehat{\cal E}_1
& \approx &
\E + \frac{1}{2 m_0 c^2} \beta \O^2 - \frac{1}{8 m_0^2 c^4}
\left[\O ,
\left(\left[\O , \E \right] +  \i \hbar \dO \right) \right] \nn \\
& & \quad
- \frac{1}{8 m_0^3 c^6} \beta \O^4 \nn \\
\O_1
& \approx & \frac{\beta}{2 m_0 c^2}
\left(\left[\O , \E \right] + \i \hbar \dO \right)
- \frac{1}{3 m_0^2 c^4} \O^3\,,
\eea
with $\E_1$ and $\O_1$ obeying the relations
$\beta \widehat{\cal O}_1 = - \widehat{\cal O}_1 \beta$ and
$\beta \widehat{\cal E}_1 = \widehat{\cal E}_1 \beta$ exactly like
$\E$ and $\O$.  It is seen that while the term $\O$ in $\widehat{H}_D$
is of order zero with respect to the expansion parameter $1/{m_0 c^2}$
({\em i.e.}, $\O$ $=$ $O \left(\left( 1/{m_0 c^2} \right)^0 \right)$
the odd part of $\widehat{H}_D^{(1)} $, namely $\O_1$, contains only
terms of order $1/{m_0 c^2}$ and higher powers of $1/{m_0 c^2}$
({\em i.e.}, $\O_1 = O \left(\left(1/{m_0 c^2}\right) \right)$).

To reduce the strength of the odd terms further in the transformed
Hamiltonian a second Foldy-Wouthuysen transformation is applied with
the same prescription:
%12
\bea
\Psi^{(2)}
& = &
e^{\i \widehat{S}_2} \Psi^{(1)} \,, \nn \\
\qquad \widehat{S}_2
& = &
- \frac{\i \beta \widehat{\cal O}_1}{2 m_0 c^2} \nn \\
& = &
- \frac{\i \beta}{2 m_0 c^2} \left[
\frac{\beta}{2 m_0 c^2}
\left( \left[\O , \E \right] + \i \hbar \dO \right)
- \frac{1}{3 m_0^2 c^4} \O^3 \right]\,. %\nn \\
\eea
After this transformation,
%13
\bea
\i \hbar \frac{\partial}{\partial t} \Psi^{(2)}
& = &
\widehat{H}_D^{(2)} \Psi^{(2)}\,, \quad
\widehat{H}_D^{(2)}
=
{m_0 c^2} \beta + \widehat{\cal E}_2 + \widehat{\cal O}_2 \nn \\
\widehat{\cal E}_2
& \approx &
\E_1\,, \quad
\O_2 \approx \frac{\beta}{2 m_0 c^2}
\left(\left[\O_1 , \E_1 \right] + \i \hbar
\frac{\partial \O_1}{\partial t} \right)\,, %\nn \\
%& &
\eea
where, now, $\O_2 = O \left(\left(1/{m_0 c^2}\right)^2 \right)$.
After the third transformation
%14
\beq
\Psi^{(3)} = e^{\i \widehat{S}_3}\,\Psi^{(2)}, \qquad
\widehat{S}_3 =
- \frac{ \i \beta \widehat{\cal O}_2}{2 m_0 c^2}\,,
\eeq
we have
%15
\bea
\i \hbar \frac{\partial}{\partial t} \Psi^{(3)}
& = &
\widehat{H}_D^{(3)} \Psi^{(3)}\,, \quad
\widehat{H}_D^{(3)}
=
{m_0 c^2} \beta + \widehat{\cal E}_3 + \widehat{\cal O}_3 \nn \\
\widehat{\cal E}_3 & \approx & \E_2 \approx \E_1\,, \quad
\O_3 \approx \frac{\beta}{2 m_0 c^2} \left(\left[\O_2 , \E_2 \right]
+ \i \hbar \frac{\partial \O_2}{\partial t} \right)\,, %\nn \\
%& &
\eea
where $\O_3 = O \left( \left( 1/{m_0 c^2} \right)^3 \right)$. So,
neglecting $\O_3$,
%16
\bea
\widehat{H}_D^{(3)}
& \approx &
{m_0 c^2} \beta + \widehat{\cal E} +
\frac{1}{2 m_0 c^2} \beta \widehat{\cal O}^2 \nn \\
& & \quad
- \frac{1}{8 m_0^2 c^4} \left[\O , \left( \left[\O , \E \right]
+ \i \hbar \frac{\partial \O }{\partial t} \right) \right] \nn \\
& & \quad
-
\frac{1}{8 m_0^3 c^6} \beta
\left\{
\O^4
+
\left(\left[\O , \E \right] + \i \hbar
\frac{\partial \O }{\partial t} \right)^2
\right\}
\label{A-FW-FOUR}
\eea
It may be noted that starting with the second transformation
successive $(\E , \O)$ pairs can be obtained recursively using the
rule
%17
\bea
\E_j & = & \E_1 \left(\E \rightarrow \E_{j-1} ,
\O \rightarrow \O_{j-1} \right) \nn \\
\O_j
& = &
\O_1 \left(\E \rightarrow \E_{j-1} ,
\O \rightarrow \O_{j-1} \right)\,, \quad j > 1\,,
\eea
and retaining only the relevant terms of desired order at each step.

With $\widehat{\cal E} = q \phi$ and
$\widehat{\cal O} = c \Al \cdot \widehat{\vpi}$, the final reduced
Hamiltonian~(\ref{A-FW-FOUR}) is, to the order calculated,
%17
\bea
\widehat{H}_D^{(3)}
& = &
\beta \left({m_0 c^2} + \frac{\widehat{\pi}^2}{2 m_0}
- \frac{\widehat{p}^4}{8 m_0^3 c^6} \right) + q \phi
- \frac{q \hbar}{2 m_0 c} \beta \Vsig \cdot \B \nn \\
& & \quad
- \frac{\i q {\hbar}^2}{8 m_0^2 c^2} \Vsig \cdot
{\rm curl}\,{\mbox{\boldmath $E$}}
- \frac{q{\hbar}}{4 m_0^2 c^2} \Vsig \cdot
{\mbox{\boldmath $E$}} \times \widehat{\p} \nn \\
& & \quad
- \frac{q{\hbar}^2}{8 m_0^2 c^2}
{\rm div}{\mbox{\boldmath $E$}}\,,
\eea
with the individual terms having direct physical interpretations. The
terms in the first parenthesis result from the expansion of
$\sqrt{m_0^2 c^4 + c^2 \widehat{\pi}^2}$
showing the effect of the relativistic mass increase. The second and
third terms are the electrostatic and magnetic dipole energies. The
next two terms, taken together (for hermiticity), contain the
spin-orbit interaction. The last term, the so-called Darwin term,
is attributed to the {\em zitterbewegung} (trembling motion) of the
Dirac particle: because of the rapid coordinate fluctuations over
distances of the order of the Compton wavelength ($2 \pi \hbar /m_0 c$)
the particle sees a somewhat smeared out electric potential.

It is clear that the Foldy-Wouthuysen transformation technique expands
the Dirac Hamiltonian as a power series in the parameter
$1/{m_0 c^2}$ enabling the use of a systematic approximation
procedure for studying the deviations from the nonrelativistic
situation.  We note the analogy between the nonrelativistic
particle dynamics and paraxial optics:

\begin{center}
{\bf The Analogy}
\end{center}
\begin{tabular}{ll}
{\bf Standard Dirac Equation} ~~~~~~~ & {\bf Beam Optical Form} \\
$m_0 c^2 \beta + \E_D + \O_D$ & $- n_0 \sigma_z + \E + \O$ \\
$m_0 c^2$ & $- n_0$ \\
Positive Energy & Forward Propagation \\
Nonrelativistic, $\left| \widehat{\vpi} \right| \ll m_0 c$ &
Paraxial Beam, $\left|\hatp_\perp \right| \ll n_0$ \\
%Deviations from nonrelativistic situation & Deviations from paraxial condition \\
Non relativistic Motion  & Paraxial Behavior \\
~~ + Relativistic Corrections & ~~ + Aberration Corrections \\
\end{tabular}

\bigskip

Noting the above analogy, the idea of Foldy-Wouthuysen form of the
Dirac theory has been adopted to study the paraxial optics and
deviations from it by first casting the Maxwell equations in a spinor
form resembling exactly the Dirac equation~(\ref{A-FW-1}, \ref{A-FW-2})
in all respects: {\em i.e}., a multicomponent $\Psi$ having the upper
half of its components large compared to the lower components and the
Hamiltonian having an even part $(\E)$, an odd part $(\O)$, a suitable
expansion parameter, ($|\hatp_\perp|/{n_0} \ll 1$) characterizing the
dominant forward
propagation and a leading term with a $\beta$ coefficient commuting
with $\E$ and anticommuting with $\O$.  The additional feature of
our formalism is to return finally to the original representation
after making an extra approximation, dropping $\beta$ from the final
reduced optical Hamiltonian, taking into account the fact that we are
primarily interested only in the forward-propagating beam.

\newpage

\section*{}
\addcontentsline{toc}{section}
{Appendix C. \\
The Magnus Formula}

%MAGNUS
\renewcommand{\theequation}{C.{\arabic{equation}}}
\setcounter{equation}{0}

\begin{center}

{\Large\bf
Appendix-C \\
The Magnus Formula
} \\

\end{center}

The Magnus formula is the continuous analogue of the famous
Baker-Campbell-Hausdorff (BCH) formula
%01
\beq
\e ^{{\hat A}} \e ^{{\hat B}} =
\e ^{ {\hat A} +  {\hat B} +
\half [ {\hat A}, {\hat B}] + \frac{1}{12} \left\{
[ [ {\hat A}, {\hat A}], {\hat B} ] +
[ [ {\hat A}, {\hat B}], {\hat B} ] \right\} + \ldots }\,.
\label{BCH}
\eeq
Let it be required to solve the differential equation
%02
\beq
\frac{\partial}{\partial t} u(t) = {\hat A} (t) u(t)
\label{Magnus-2}
\eeq
to get $u(T)$ at $T > t_0$, given the value of $u (t_0)$; the
operator ${\hat A}$ can represent any linear operation.  For an
infinitesimal $ \Delta t$, we can write
%03
\beq
u (t_0 + \Delta t) = e^{\Delta t {\hat A}(t_0)} u (t_0).
\eeq
Iterating this solution we have
%04
\bea
u(t_0 + 2 \Delta t) & = & \e ^ {\Delta t {\hat A}(t_0 + \Delta t)}
\e ^ {\Delta t {\hat A}(t_0 )} u(t_0)
\nn \\
u(t_0 + 3 \Delta t) & = & \e ^ {\Delta t {\hat A}(t_0 + 2 \Delta t)}
\e ^ {\Delta t {\hat A}(t_0 + \Delta t)}
\e ^ {\Delta t {\hat A}(t_0 )} u(t_0)
\nn \\
&   & \ldots \quad {\rm and\ so\ on}.
\eea
If $ T = t_0 + N \Delta t$ we would have
%05
\beq
u(T) = \left\{ \prod_{n =0}^{N-1}
e ^ {\Delta t {\hat A}(t_0 + n \Delta t)} \right\} u(t_0)\,.
\label{pi}
\eeq
Thus, $u(T)$ is given by computing the product in~(\ref{pi})
using
successively the BCH-formula~(\ref{BCH}) and considering the limit $\Delta t
\longrightarrow 0, N \longrightarrow \infty $ such that
$N \Delta t = T -t_0 $. The resulting expression is the Magnus
formula~(Magnus,~\cite{Magnus})~:
%06
\bea
u(T)
& = &
\widehat{\cal T}(T, t_0) u(t_0)
\nn \\
{\cal T}(T, t_0)
& = &
\exp \left\{\int_{t_0}^{T} d t_1\,{\hat A}(t_1) \right.  \nn \\
& & \
+ \half \int_{t_0}^{T} d t_2
\int_{t_0}^{t_2} d t_1\,\left[{\widehat A}(t_2), {\hat A}(t_1) \right]
\nn \\
&  & \ + \frac{1}{6}
\int_{t_0}^{T} d t_3 \int_{t_0}^{t_3} d t_2 \int_{t_0}^{t_2} d t_1\,\left(
\left[ \left[ {\hat A}(t_3), {\hat A}(t_2)\right], {\hat A}(t_1) \right] \right.
\nn \\
&  & \left. \phantom{\int_{t_0}^T} {\left. \quad \quad  + \left[ \left[ {\hat
A}(t_1), {\hat A}(t_2) \right], {\hat A}(t_3) \right] \right)} +\,\ldots
\right\}\,.
\label{Magnus-6}
\eea

To see how the equation~(\ref{Magnus-6}) is obtained let us substitute
the assumed form of the solution,
$u (t) = \ct \left(t , t_0 \right) u \left(t_0 \right)$,
in~(\ref{Magnus-2}).  Then, it is seen that
$\widehat{{\cal T}} (t, t_0)$ obeys the equation
%07
\beq
\frac{\partial }{\partial t} \widehat{{\cal T}} (t, t_0) =
{\widehat A} (t) {\cal T} (t, t_0),
\qquad {\widehat{\cal T}} (t_0, t_0) = {\widehat I}\,.
\label{Magnus-7}
\eeq
Introducing an iteration parameter $\lambda$ in~(\ref{Magnus-7}), let
%08-09
\bea
\frac{\partial}{\partial t} \widehat{{\cal T}}(t, t_0; \lambda)
& = &
\lambda {\widehat A}(t) \widehat{{\cal T}}(t, t_0; \lambda)\,,
\label{Magnus-8}
\\
\widehat{{\cal T}}(t_0, t_0; \lambda)
& = &
{\widehat I}\,, \quad
\widehat{{\cal T}}(t, t_0; 1) = \widehat{{\cal T}}(t, t_0)\,.
\label{Magnus-9}
\eea
Assume a solution of~(\ref{Magnus-8}) to be of the form
%10
\beq
\widehat{{\cal T}}(t, t_0; \lambda)
= \e ^{{\Omega} (t, t_0; \lambda)}
\eeq
with
%11
\beq
{\Omega}(t, t_0; \lambda)
= \sum_{n =1}^{\infty} {\lambda} ^n
\Delta_n (t, t_0), \quad \Delta_n (t_0, t_0) = 0 \quad
{\rm for \ all \ } n\,.
\label{Magnus-11}
\eeq
Now, using the identity (see, Wilcox,~\cite{Wilcox})
%12
\beq
\frac{\partial}{\partial t} \e ^{ {\Omega}(t, t_0; \lambda)} =
\left\{
\int_{0}^{1} ds e ^{s \Omega (t, t_0; \lambda)}
\frac{\partial}{\partial t} {\Omega}(t, t_0; \lambda)
\e ^{ - s \Omega (t, t_0; \lambda)}  \right\} \e ^{{\Omega}(t, \lambda)}\,,
\eeq
one has
%13
\beq
\int_{0}^{1} ds e ^{s \Omega (t, t_0; \lambda)}
\frac{\partial}{\partial t} {\Omega}(t, t_0; \lambda)
e ^{ - s \Omega (t, t_0; \lambda)} =
\lambda \widehat{A} (t)\,.
\eeq
Substituting in~(A13) the series expression for
$\Omega (t, t_0; \lambda)$~(\ref{Magnus-11}), expanding the
left hand side using
the first identity in~(C8), integrating and equating the
coefficients of $ \lambda ^j$ on both
sides, we get, recursively, the equations for
$ \Delta_1 (t, t_0)$, $\Delta_2 (t, t_0), \ldots \,,$ etc. For $j = 1$
%14
\beq
\frac{\partial}{\partial t} \Delta_1 (t, t_0) = {\hat A}(t),
\qquad \Delta_1 (t_0, t_0) = 0
\eeq
and hence
%15
\beq
\Delta_1 (t, t_0) =  \int_{t_0}^{t} d t_1 \widehat{A} (t_1)\,.
\eeq
For $j=2$
%16
\beq
\frac{\partial}{\partial t} \Delta_2 (t, t_0) +
\half \left[ \Delta_1 (t, t_0)\,,\,\frac{\partial}{\partial t}
\Delta_1 (t, t_0) \right] =  0\,,
\qquad \Delta_2 (t_0, t_0) = 0
\eeq
and hence
%17
\beq
\Delta_2 (t, t_0) =  \half
\int_{t_0}^{t} d t_2
\int_{t_0}^{t_2} d t_1 \left[\widehat{A} (t_2)\,,\,
\widehat{A} (t_1) \right].
\eeq
Similarly,
%18
\bea
\Delta_3 (t, t_0) & = & \frac{1}{6}
\int_{t_0}^{t} d t_1 \int_{t_0}^{t_1} d t_2 \int_{t_0}^{t_2} d t_3\,\left\{
\left[ \left[ {\hat A}(t_1)\,,\,{\hat A}(t_2) \right]\,,\,{\hat A}(t_3) \right]
\right.
\nn \\
&   & \quad \quad \left. + \left[ \left[ {\hat A}(t_3)\,,\,{\hat A}(t_2)
\right]\,,\,{\hat A}(t_1) \right] \right\}\,.
\eea
Then, the Magnus formula in~(\ref{Magnus-6}) follows
from~(\ref{Magnus-9})-(\ref{Magnus-11}).
Equation~\ref{T-Magnus} we have, in the context of $z$-evolution
follows
from the above discussion with the identification
$t \longrightarrow z$, $t_0 \longrightarrow z^{(1)}$,
$T \longrightarrow z^{(2)}$
and $ {\hat A}(t) \longrightarrow - \ih \ho (z)$.

For more details on the exponential solutions of linear differential
equations, related operator techniques and applications to
physical problems the reader is referred to Wilcox~\cite{Wilcox},
Bellman and Vasudevan~\cite{BV}, Dattoli {\em et al.}~\cite{DRT},
and references therein.

\newpage

\section*{}
\addcontentsline{toc}{section}
{Appendix D. \\
Analogies between light optics and charged-particle optics:
Recent Developments
}

%ANALOGY
\renewcommand{\theequation}{D.{\arabic{equation}}}
\renewcommand{\thesection}{D.{\arabic{section}}}
\renewcommand{\thesubsection}{D.{\arabic{subsection}}}
\setcounter{subsection}{0}
\setcounter{equation}{0}

\begin{center}

{\Large\bf
Appendix-D \\
Analogies between light optics and charged-particle optics:
Recent Developments
} \\

\end{center}

Historically, variational principles have played a fundamental role
in the evolution of mathematical models in classical physics, and many
equations can be derived by using them.  Here the relevant examples are
Fermat's principle in optics and Maupertuis' principle in mechanics.
The beginning of the analogy between geometrical optics and mechanics
is usually attributed to Descartes (1637), but actually it can traced
back to Ibn Al-Haitham Alhazen (0965-1037)~\cite{Ambrosini}.  The
analogy between the trajectory of material particles in potential
fields and the path of light rays in media with continuously variable
refractive index was formalized by Hamilton in 1833.  The Hamiltonian
analogy lead to the development of electron optics in 1920s, when Busch
derived the focusing action and a lens-like action of the axially
symmetric magnetic field using the methodology of geometrical optics.
Around the same time Louis de Broglie associated his now famous
wavelength to moving particles.  Schr\"{o}dinger extended the analogy
by passing from geometrical optics to wave optics through his wave
equation incorporating the de Broglie wavelength.  This analogy played
a fundamental role in the early development of quantum mechanics.  The
analogy, on the other hand, lead to the development of practical
electron optics and one of the early inventions was the electron
microscope by Ernst Ruska.  A detailed account of Hamilton's analogy
is available in~\cite{Hawkes}-\cite{Forbes}.

Until very recently, it was possible to see this analogy only between
the geometrical-optic and classical prescriptions of electron optics.
The reasons being that, the quantum theories of charged-particle beam
optics have been under development only for about a
decade~\cite{JSSM}-~\cite{JK} with the very expected feature of
wavelength-dependent effects, which have no analogue in the traditional
descriptions of light beam optics.  With the  current development of
the non-traditional prescriptions of Helmholtz optics~\cite{KJS-1} and
the matrix formulation of Maxwell optics, accompanied with
wavelength-dependent effects, it is seen that the analogy between the
two systems persists.  The non-traditional prescription of Helmholtz
optics is in close analogy with the quantum theory of charged-particle
beam optics based on the Klein-Gordon equation.  The matrix formulation
of Maxwell optics is in close analogy with the quantum theory of
charged-particle beam optics based on the Dirac equation.  This analogy
is summarized in the table of Hamiltonians.  In this short note it is
difficult to present the derivation of the various Hamiltonians from
the quantum theory of charged-particle beam optics, which are all
available in the references.  We shall briefly consider an outline of
the quantum prescriptions and the non-traditional prescriptions
respectively.  A complete coverage to the new field of  {\em Quantum
Aspects of Beam Physics} ({\bf QABP}), can be found in the proceedings
of the series of meetings under the same name~\cite{QABP}.

\subsection{Quantum Formalism of Charged-Particle Beam Optics}
The classical treatment of charged-particle beam optics has been
extremely successful in the designing and working of numerous optical
devices, from electron microscopes to very large particle accelerators.
It is natural, however to look for a prescription based on the quantum
theory, since any physical system is quantum mechanical at the
fundamental level!  Such a
prescription is sure to explain the grand success of the classical
theories and may also help get a deeper understanding and to lead to
better designing of charged-particle beam devices.

The starting point to obtain a quantum prescription of charged particle
beam optics is to build a theory based on the basic equations
(Schr\"{o}dinger, Klein-Gordon, Dirac) of quantum mechanics appropriate
to the situation under study.  In order to analyze the evolution of the
beam parameters of the various individual beam optical elements
(quadrupoles, bending magnets,~$\cdots$) along the optic axis of the
system, the first step is to start with the basic time-dependent
equations of quantum mechanics and then obtain an equation of the form
%01
\begin{equation}
\i \hbar \frac{\partial }{\partial s} \psi \left(x , y ;\, s \right)
=
\widehat{\cal H} \left(x , y ;\, s \right)
\psi \left(x , y ;\, s \right)\,,
\label{BOE}
\end{equation}
where $(x , y ;\, s)$ constitute a curvilinear coordinate
system, adapted to the geometry of the system.  Eq.~(\ref{BOE}) is
the basic equation in the quantum formalism, called as the
{\em beam-optical equation}; ${\cal H}$ and $\psi$ as the
{\em beam-optical Hamiltonian} and the {\em beam wavefunction}
respectively.  The second step requires obtaining a relationship
between any relevant observable $\{\langle O \rangle (s) \}$ at the
transverse-plane at $s$ and the observable
$\{\langle O \rangle (s_{\rm in}) \}$
at the transverse plane at $s _{\rm in}$, where $s _{\rm in}$ is some
input reference point.  This is achieved by the integration of the
beam-optical equation in~(\ref{BOE})
%02
\begin{eqnarray}
\psi \left(x , y ; s \right) & = &
\widehat{U} \left(s , s_{\rm in} \right)
\psi \left(x , y ; s_{\rm in} \right)\,,
\label{BOI}
\end{eqnarray}
which gives the required transfer maps
%03
\begin{eqnarray}
\left\langle O \right\rangle \left(s_{\rm in} \right)
\longrightarrow
\left\langle O \right\rangle \left(s \right)
& = &
\left\langle \psi \left(x , y ; s \right)
\left| O \right|
\psi \left(x , y ; s \right) \right\rangle\,, \nn \\
& = &
\left\langle \psi \left(x , y ; s_{\rm in} \right)
\left| \widehat{U}^{\dagger} O \widehat{U} \right|
\psi \left(x , y ; s_{\rm in} \right) \right\rangle\,.
\label{BOM}
\end{eqnarray}

The two-step algorithm stated above gives an over-simplified picture of
the quantum formalism.  There are several crucial points to be noted.
The first-step in the algorithm of obtaining the beam-optical equation
is not to be treated as a mere transformation which eliminates $t$ in
preference to a variable $s$ along the optic axis.  A clever set of
transforms are required which not only eliminate the variable $t$ in
preference to $s$ but also give us the $s$-dependent equation which has
a close physical and mathematical analogy with the original
$t$-dependent equation of standard time-dependent quantum mechanics.
The imposition of this stringent requirement on the construction of the
beam-optical equation ensures the execution of the second-step of the
algorithm.  The beam-optical equation is such that all the required
rich machinery of quantum mechanics becomes applicable to the
computation of the transfer maps that characterize the optical system.
This describes the essential scheme of obtaining the quantum formalism.
The rest is mostly mathematical detail which is inbuilt in the powerful
algebraic machinery of the algorithm, accompanied with some reasonable
assumptions and approximations dictated by the physical considerations.
The nature of these approximations can be best summarized in the optical
terminology as a systematic procedure of expanding the beam optical
Hamiltonian in a power series of $|{\widehat{\vpi}_\perp}/{p_0}|$,
where $p_0$ is the design (or average) momentum of beam particles
moving predominantly along the direction of the optic axis and
$\widehat{\vpi}_\perp$ is the small transverse kinetic momentum.  The
leading order approximation along with
$|{\widehat{\vpi}_\perp}/{p_0}| \ll 1$, constitutes the paraxial or
ideal behaviour and higher order terms in the expansion give rise to
the nonlinear or aberrating behaviour.
It is seen that the paraxial and aberrating behaviour get modified by
the quantum contributions which are in powers of the de Broglie
wavelength ($\bar{\lambda}_0 = {\hbar}/{p_0}$).  The classical limit
of the quantum formalism reproduces the well known Lie algebraic
formalism~\cite{Lie} of charged-particle beam optics.

\subsection{Light Optics: Various Prescriptions}
The traditional scalar wave theory of optics (including aberrations to
all orders) is based on the beam-optical Hamiltonian derived by using
Fermat's principle.  This approach is purely geometrical and works
adequately in the scalar regime.  The other approach is based on the
{\em square-root} of the Helmholtz operator, which is derived from the
Maxwell equations~\cite{Lie}.  This approach works to all orders and
the resulting expansion is no different from the one obtained using
the geometrical approach of Fermat's principle.  As for the
polarization: a systematic procedure for the passage from scalar to
vector wave optics to handle paraxial beam propagation problems,
completely taking into account the way in which the Maxwell equations
couple the spatial variation and polarization of light waves, has been
formulated by analyzing the basic Poincar\'{e} invariance of the system,
and this procedure has been successfully used to clarify several issues
in Maxwell optics~\cite{MSS-1,SSM-1,SSM-2}.

The two-step algorithm used in the construction of the quantum theories
of charged-particle beam optics is very much applicable in light optics!
But there are some very significant conceptual differences to be born
in mind.  When going beyond Fermat's principle the whole of optics
is completely governed by the Maxwell equations, and there are no other
equations, unlike in quantum mechanics, where there are separate
equations for, spin-$1/2$, spin-$1$, $\cdots$.

Maxwell's equations are linear (in time and space derivatives) but
coupled in the fields.  The decoupling leads to the Helmholtz equation
which is quadratic in derivatives.  In the specific context of beam
optics, purely from a calculational point of view, the starting
equations are the Helmholtz equation governing scalar optics and for a
more complete prescription one uses the full set of Maxwell equations,
leading to vector optics.  In the context of the two-step algorithm,
the Helmholtz equation and the Maxwell equations in a matrix
representation can be treated as the `basic' equations, analogue of
the basic equations of quantum mechanics.  This works perfectly fine
from a calculational point of view in the scheme of the algorithm we
have.

Exploiting the similarity between the Helmholtz wave equation and the
Klein-Gordon equation, the former is linearized using
the Feshbach-Villars procedure used for the linearization of the
Klein-Gordon equation.  Then the Foldy-Wouthuysen iterative
diagonalization technique is applied to obtain a Hamiltonian description
for a system with varying refractive index.  This technique is an
alternative to the conventional method of series expansion of the
radical.  Besides reproducing all the traditional quasiparaxial terms,
this method leads to additional terms, which are dependent on the
wavelength, in the optical Hamiltonian.
This is the non-traditional prescription of scalar optics.

The Maxwell equations are cast into an exact matrix form taking into
account the spatial and temporal variations of the permittivity and
permeability.  The derived  representation using $8 \times 8$ matrices
has a close algebraic analogy with the Dirac equation, enabling the use
of the rich machinery of the Dirac electron theory.  The beam optical
Hamiltonian derived from this representation reproduces the
Hamiltonians obtained in the traditional prescription along with
wavelength-dependent matrix terms, which we have named as the
{\em polarization terms}~\cite{Khan-Maxwell}.  These polarization terms
are algebraically very similar to the spin terms in the Dirac electron
theory and the spin-precession terms in the beam-optical version of the
Thomas-BMT equation\cite{CJKP-1}.  The  matrix formulation provides a
unified treatment of beam optics and light polarization.  Some well
known results of light polarization are obtained as a paraxial limit of
the matrix formulation~\cite{MSS-1,SSM-1,SSM-2}.
The traditional beam optics is completely obtained from our approach
in the limit of small wavelength, $\LAMBDA \longrightarrow 0$, which
we call as the traditional limit of our formalisms.  This is analogous
to the classical limit obtained by taking $\hbar \longrightarrow 0$,
in the quantum prescriptions.

From the Hamiltonians in the Table we make the following observations:
The classical/traditional Hamiltonians of particle/light optics are
modified by wavelength-dependent contributions in the
quantum/non-traditional prescriptions respectively.  The algebraic
forms of these modifications in each row is very similar.  This should
not come as a big surprise.  The starting equations have one-to-one
algebraic correspondence: Helmholtz $\leftrightarrow$ Klein-Gordon;
Matrix form of Maxwell $\leftrightarrow$ Dirac equation.  Lastly, the
de Broglie wavelength, $\bar{\lambda}_0$, and $\LAMBDA$ have an
analogous status, and the classical/traditional limit is obtained by
taking $\bar{\lambda}_0 \longrightarrow 0$ and
$\LAMBDA \longrightarrow 0$ respectively.  The parallel of the
analogies between the two systems is sure to provide us with more
insights.

\newpage

\section*{}
\addcontentsline{toc}{section}
{Table A. \\
Hamiltonians in Different Prescriptions}

%TABLE-A

\begin{center}
{\Large\bf
Table A. \\
Hamiltonians in Different Prescriptions
} \\
\end{center}

{\small
\noindent
The following are the Hamiltonians, in the different prescriptions
of light beam optics and charged-particle beam optics for magnetic
systems.  $\widehat{H}_{0\,, p}$ are the paraxial Hamiltonians, with
lowest order wavelength-dependent contributions.

\noindent
\begin{tabular*}{6.0in}[t]{@{\extracolsep{\fill}}|ll|}
\hline
& \\
\parbox[t]{2.5in}{
\parbox[t]{2.5in}{
{\large\bf Light Beam Optics}
}}
&
\parbox[t]{3.0in}{
{\large\bf Charged-Particle Beam Optics}
} \\
& \\
\hline
& \\
\parbox[t]{2.5in}{
{\bf Fermat's Principle} \\

$
{\cal H}
=
- \left\{n^2 (\r) - \p_\perp^2 \right\}^{1/2}
$
}

&

\parbox[t]{2.5in}{
{\bf Maupertuis' Principle} \\

$
{\cal H}
=
- \left\{p_0^2 - {\vpi}_\perp^2 \right\}^{1/2} - q A_z
$
} \\
& \\
\hline
& \\
\parbox[t]{2.5in}{

{\bf Non-Traditional Helmholtz} \\

$
\widehat{H}_{0\,, p} = \\
- n (\r)
+ \frac{1}{2 n_0} \hatp_{\perp}^2 \\
- \frac{\i \LAMBDA}{16 n_0^3}
\left[\hatp_\perp^2 , \ddz n (\r) \right]
$

}

&

\parbox[t]{2.5in}{
{\bf Klein-Gordon Formalism} \\

$
\widehat{H}_{0\,, p} = \\
- p_0 - q A_z + \frac{1}{2 p_0} \widehat{\vpi}_\perp^2 \\
+ \frac{\i \hbar}{16 p_0^4}
\left[\widehat{\vpi}_\perp^2 \,, \ddz \widehat{\vpi}_\perp^2 \right]
$
} \\
& \\
\hline
& \\

\parbox[t]{2.5in}{

{\bf Maxwell, Matrix} \\

$
\widehat{H}_{0\,, p} = \\
- n (\r) + \frac{1}{2 n_0} \hatp_\perp^2 \\
- \i \LAMBDA \beta
{\mbox{\boldmath $\Sigma$}} \cdot {\mbox{\boldmath $u$}} \\
+
\frac{1}{2 n_0} \LAMBDA^2 w^2 \beta
$
}

&

\parbox[t]{2.5in}{
{\bf Dirac Formalism} \\

$
\widehat{H}_{0\,, p} = \\
- p_0 - q A_z  + \frac{1}{2 p_0} \widehat{\vpi}_\perp^2 \\
- \frac{\hbar}{2 p_0}
\left\{\mu \gamma
{\mbox{\boldmath $\Sigma$}}_\perp \cdot \B_\perp
+
\left(q + \mu \right) \Sigma_z B_z \right\} \\
+ \i \frac{\hbar}{m_0 c} \epsilon B_z
$
} \\
& \\
\hline
\end{tabular*}

\bigskip

\noindent
{\bf Notation}

\noindent
\begin{tabular*}{6.0in}[t]{@{\extracolsep{\fill}}ll}
\parbox[t]{2.5in}{
\parbox[t]{2.5in}{
$
{\rm Refractive ~ Index}, ~
n (\r) = c \sqrt{\epsilon (\r) \mu (\r)} \\
{\rm Resistance}, ~
h (\r) = \sqrt{{\mu (\r)}/{\epsilon (\r)}} \\
{\mbox{\boldmath $u$}} (\r)
=
- \frac{1}{2 n (\r)} \Nab n (\r) \\
{\mbox{\boldmath $w$}} (\r)
=
\frac{1}{2 h (\r)} \Nab h (\r) \\
$
${\mbox{\boldmath $\Sigma$}}$ and $\beta$ are the Dirac matrices.
}}

&

\parbox[t]{2.5in}{
$
\widehat{\vpi}_\perp
= {\widehat{\mbox{\boldmath $p$}}}_\perp
- q {\mbox{\boldmath $A$}}_\perp \\
\mu_a ~ {\rm anomalous ~ magnetic ~ moment}. \\
\epsilon_a ~ {\rm anomalous ~ electric ~ moment}. \\
\mu = {2 m_0 \mu_a}/{\hbar}\,, ~~~~
\epsilon = {2 m_0 \epsilon_a}/{\hbar} \\
\gamma = {E}/{m_0 c^2}
$
}
\end{tabular*}

}

\newpage

%\end{document}

\end{document}